\documentclass{sig-alternate-05-2015}

\usepackage{amssymb}
\usepackage{amsmath}
\usepackage{color}
\usepackage{times}
\usepackage{cite}
\usepackage{enumitem}
\usepackage{arydshln}
\usepackage{url}
\usepackage{algorithm}
\usepackage{algorithmic}
\usepackage{flushend}
\usepackage{graphicx}

\usepackage{enumitem}
\setitemize{noitemsep,topsep=0pt,parsep=0pt,partopsep=0pt}

\newcommand{\hindent}[1][1]{\hspace{#1\algorithmicindent}}

\title{HDBSCAN: Density based Clustering over Location Based Services}
\numberofauthors{1}
\author{
	Md Farhadur Rahman$^\ddag$,
	Weimo Liu$^\dag$,
	Saad Bin Suhaim$^\dag$,
	Saravanan Thirumuruganathan$^\ddag$,\\
	Nan Zhang$^\dag$,
	Gautam Das$^\ddag$\\
	\affaddr{ University of Texas at Arlington$^\ddag$, The George Washington University$^\dag$}\\
}

\begin{document}
\setcopyright{acmcopyright}

\maketitle

\begin{abstract}
Location Based Services (LBS) have become extremely popular and used by millions of users. Popular LBS run the entire gamut from mapping services (such as Google Maps) to restaurants (such as Yelp) and real-estate (such as Redfin). The public query interfaces of LBS can be abstractly modeled as a $k$NN interface over a database of two dimensional points: given an arbitrary query point, the system returns the $k$ points in the database that are nearest to the query point. Often, $k$ is set to a small value such as 20 or 50. In this paper, we consider the novel problem of enabling density based clustering over an LBS with only a limited, $k$NN query interface. Due to the query rate limits imposed by LBS, even retrieving every tuple once is infeasible. Hence, we seek to construct a cluster assignment function $f(\cdot)$ by issuing a small number of kNN queries, such that for any given tuple $t$ in the database which may or may not have been accessed, $f(\cdot)$ outputs the cluster assignment of $t$ with high accuracy. We conduct a comprehensive set of experiments over benchmark datasets and popular real-world LBS such as Yahoo! Flickr, Zillow, Redfin and Google Maps. 
\end{abstract}

\section{Introduction}
\label{sec:intro}

\smallskip\noindent
{\bf Location Based Services (LBS):} LBS have become extremely popular and used by millions of users. They are an essential data source for geospatial and commercial information such as Points-of-Interest (POIs), restaurants, real-estate, user geo-distribution etc. Popular LBS run the entire gamut from mapping services (such as Google Maps) to restaurants reviews (such as Yelp) and real-estate (such as Redfin) etc. In addition, LBS related features have been widely integrated into other popular platforms such as WeChat, Sina Weibo etc. 
Each LBS has a backend database with tuples representing POIs or users, and attributes capturing their geographical coordinates (e.g., latitude and longitude) along with other information such as POI name, review ratings, etc.
What the LBS reveals to the public is typically a ``local view'' of the database, such as a $k$NN query interface or API which, upon given an arbitrary query location and a user-specified selection condition, returns a small number of (top-$k$) tuples in the database that, among those matching the selection condition, are closest (geographically) to the query location.  Additionally, the LBS also impose a limit on the maximum number of queries one can issue per user/IP.

\smallskip\noindent
{\bf Problem Definition and Motivation:}
The backend databases of LBS are extremely important data sources, since they contain a gold mine of information for understanding POI quality, user behavior, and other big picture information about the underlying data. In particular, since the LBS data providers have complete data access (i.e. ``global view'' of the data), they can use a plethora of well-known spatial data mining algorithms (e.g., {\em density-based clustering} algorithms such as DBSCAN \cite{ester1996density} and DENCLUE \cite{hinneburg1998efficient}) to gain insights into the spatial distributions of the data. 

In this paper we investigate the following novel problem:  Is it possible to develop density-based clustering algorithms over LBS data that {\em only have local-view access to the data, such as a $k$NN query interface}? 
Clustering is one of the key techniques in spatial data mining and such algorithms can enable big picture analytics of LBS databases by public and third-party clients, and not be limited only to the data providers themselves. 
 For example, a third party application might use our tool to identify hotspots or events that are occurring over Twitter. Or a user might use it to find interesting patterns in real-estate (by say running it over Redfin) or POIs (running it over a city in Google Maps).  
We focus on the fundamental and popular density-based clustering algorithm DBSCAN, and develop a DBSCAN-like algorithm for LBS data with only a $k$NN interface for data access. Extending our results to other density-based clustering algorithms is left for future work.  

\smallskip\noindent
{\bf Key Challenges:}
There are many challenges in developing a DBSCAN-like algorithm over LBS data with only a $k$NN query interface. The foremost challenge is that there is no direct way for a LBS client to run  DBSCAN, since the user can only formulate $k$NN search queries, and not clustering queries. Any approach that works has to be based on executing a set of $k$NN queries via the restrictive query interface, and then ``inferring'' the underlying clusters structure from the results of the queries. 

The second challenge is that a faithful implementation of  DBSCAN requires us to retrieve each tuple at least once, because the output of DBSCAN is a labeling of each tuple into the cluster class it belongs to. This is tantamount to a complete crawl of the data by making numerous queries to the $k$NN interface, which can be prohibitively expensive and may violate the rate limits or budgetary constraints imposed by the LBS interface (e.g., Google map API imposes a query rate limit of 10,000 per user per day). 

Thus, our goal here is not to identify the cluster for each tuple in the database. Instead, we aim to construct a clustering function $f(\cdot)$ by issuing a small number of $k$NN queries, such that for any given tuple $t$ in the database which {\em may or may not} have accessed, $f(\cdot)$ outputs the cluster assignment of $t$ with high accuracy. While this might seem a deviation from the traditional definition of clustering, the two problems are essentially the same. Specifically, note that by having a ``perfect'' construction of $f(\cdot)$, we would have accurately identified the number of clusters and a few tuples in each cluster. The accuracy of this $f(\cdot)$ function can also be evaluated in the same way as traditional clustering - i.e., by applying $f(\cdot)$ over all tuples in the database, we can compare the distance between its outputs and the output of a traditional clustering algorithm like DBSCAN to assess the accuracy of $f(\cdot)$.

In recent work \cite{liu2015aggrEst}, we developed an algorithm for randomly sampling tuples from an LBS database using only the $k$NN interface. A seemingly simple baseline would be to first sample the database, and then run DBSCAN over the retrieved samples. For any new point, we can simply assign the nearest cluster as its cluster assignment. This approach, while conceptually simple, also suffers from a number of issues. For example, it might mishandle arbitrarily shaped clusters - a key strength of density based clustering algorithms. Additionally, it does not have an effective way to identify outliers. Finally, it requires a substantially large sample size to get reasonably good clusters. 

In summary, our problem is substantially different from density based clustering over traditional databases. Often, they have access to entire database and their key objective is to reduce the computational cost. In contrast, our algorithms only has a local view of the database, and our objective is to perform a best-effort clustering (due to the limited data access) while minimizing the number of queries issued over the $k$NN interface. The computations that are performed at the client side are not a major limiting factor.

\smallskip\noindent
{\bf Outline of Technical Results:}
 DBSCAN is based on two key parameters $\epsilon$ and {\em minPts}, and defines a cluster as a set of points (with cardinality at least {\em minPts}) that are within a radius of $\epsilon$ from at least one other point in the cluster. It  considers a density measure equal to the number of points within a pre-determined radius $\epsilon$. Thus, a requirement for implementing DBSCAN over LBS is the ability to compute the density at any given spatial point.

One seemingly straightforward approach to simulate DBSCAN is as follows. If we partition the data space into grid cells of length $\epsilon$ on each dimension, then DBSCAN can then be almost faithfully executed as long as we can accurately estimate the number of points in each cell. Given that the key information required is whether the density of a grid cell exceeds the pre-determined density threshold {\em minPts}, our problem is reduced to estimating a Boolean indicator for each grid cell - whether the number of points falling within exceeds the given threshold {\em minPts}. This can be easily computed by issuing a $k$NN query $q$ at the center of the cell. If all $k$ returned points fall within the cell, then we return TRUE for the cell. Otherwise, we can return FALSE for not only the cell of $q$, but also all other cells completely covered within the range of the $k$ returned points. 

However, this approach can be extremely inefficient because it requires examining each cell in the grid - equivalent to visiting all points in a cluster. Hence, we need an efficient mechanism to find the
{\em boundaries} of the cluster by ``skipping over'' intermediate points that are close to each other.  Representations of these boundaries serve as our clustering function $f(\cdot)$,  since any new point can be mapped into the appropriate cluster by checking whether it lies inside or outside the cluster boundary. The problem of discovering cluster boundaries is studied in literature \cite{xia2006border,qiu2007brim}. However, like traditional clustering algorithms they are only applicable when we have full access to the database. Instead, our algorithms start by first discovering a point with high density, and then quickly traverse to the boundary of the cluster that contains that point, without having to examine all intermediate points. 

We first develop clustering algorithms for the special case of one dimensional data, and then extend the results to two (and higher) dimensions. In the 1D case, each cluster is essentially a dense segment, and our goal is to discover the boundaries of each dense segment. We develop an algorithm that starts from a dense point within a cluster and discovers the two boundary points by going to the left and right using a binary search-like process. 

In contrast to 1D case where cluster boundaries of a dense segment can be discovered by going to the left and right side from a point inside the segment, the direction that we need to follow in 2D scenario is not very clear, especially when clusters can have any arbitrary shape. In this case, we use an innovative approach of mapping the points in 2D space to 1D using a {\em space filling curve} (SFC \cite{xu2014optimality,mokbel2003analysis,moon2001analysis}), and then discover the clusters using the 1D clustering algorithm. A well designed SFC guarantees that two points close to each other in the mapped 1D space are also close together in the original 2D space. This property fits our purpose since we can skip points (by binary search) that are close to each other in mapped 1D space as they will also be potentially inside the same cluster in original 2D space. However, we must caution that points that are close to each other in 2D space might not be close in the mapped 1D space. Hence, a cluster in 2D space might split into many small clusters in 1D space. This complication can be addressed by a post-processing step of merging 1D clusters that are ``close'' to each other, eventually into 2D clusters.

The rest of the paper is organized as follows. \S\ref{sec:datamodel} introduces the LBS data model and formalizes the problem of enabling DBSCAN over LBS. \S\ref{sec:oneD} develops algorithm HDBSCAN-1D for 1D data that highlights the basic ideas of our sampling based design. \S\ref{sec:generalCase} considers the general case of clustering higher-dimensional data and develops algorithm HDBSCAN. \S\ref{sec:exp} describes the experiments over popular benchmark datasets and real-world LBS. We describe the related work in  \S\ref{sec:relWork} followed by final remarks in \S\ref{sec:finalRemarks}.

\section{Background}
\label{sec:datamodel}

\subsection{Model of LBS}
Consider a Location Based Service (LBS) over a database $D$ of $n$ tuples, each of which is labeled with a 2D location and possibly other relational attributes.  Note that the results of this paper can be directly extended to 3D (or higher dimensional) locations, as we shall discuss in \S~\ref{sec:generalCase}. Examples of such an LBS include Google Maps, Redfin, etc., where each tuple is a point of interest such as restaurant or real estate property; as well as online social networks such as WeChat where each tuple is a user.  In these examples, a tuple features not only its 2D location, but also other attributes such as restaurant rating, user gender, etc. 

LBS usually supports $k$NN queries over the database. Such a query takes as input a 2D location $q$ (e.g., $\langle$longitude, latitude$\rangle$), and returns the top-$k$ nearest tuples in $D$ to $q$ as determined by a pre-defined distance function. We consider Euclidean distance as the distance function.  For each returned tuple, the query answer includes both its location and other attributes.  Note that while the value of $k$ varies on different real-world LBS systems, it is generally at the range of 50 to 100 - e.g., Google Maps has $k = 60$, while $k = 50$ and $100$ for WeChat and Sina Weibo respectively. 
Most LBS also impose additional restrictions such as {\em query rate limit} - i.e. the maximum number of $k$NN queries that can be issued per unit of time. For example, by default Google Maps allows 10,000 location queries per day while Sina Weibo allows only 150 queries per hour. 
Given these query limits, a key goal of our algorithms is to minimize the query cost.\\

\subsection{Problem Definition}
\label{subsec:problemDefinition}

\noindent{\bf Objective of Clustering:} As discussed in the introduction, we consider in this paper how to enable clustering over an LBS that exposes nothing but the above-described, limited, $k$NN interface. An important observation here is that, with the limitation imposed by real-world LBS systems, the definition of clustering will inevitably change in our problem setting.  To understand why, note that the traditional definition of clustering is to assign a cluster ID for every tuple in the database (with the possibility of NULL ID for tuples deemed noise).  If we adopt the same goal, this means {\em accessing} each of the $n$ tuple at least once through the interface, which requires no fewer than $n/k$ queries (because each query returns at most $k$ tuples).  Given the large $n$, small $k$, and stringent query rate limit in real-world LBS systems, this query cost is often prohibitively expensive.

To address the challenge, the objective of clustering in our setting is to output a {\em cluster-assignment} function $f(\cdot)$ which, upon given a tuple $t \in D$ (which may have never been accessed by our algorithm), outputs the cluster ID of $t$.  Note that this is indeed a {\em generalization} of the original clustering definition, and the traditional practice of producing a cluster ID for each tuple can also be considered as producing a function that maps each tuple to an ID.

\vspace{1mm}
\noindent{\bf Orthogonality with Traditional Clustering Research:} Before discussing the performance measures for our clustering-over-LBS problem, it is important to make a proper distinction between the main objective of this paper and that of traditional clustering research.  In both cases, the ultimate goal is to produce $f(\cdot)$ that perfectly resembles the ground-truth cluster assignment of all tuples, e.g., as determined by human experts.  The key challenge, however, is completely different.

One can roughly partition the design of a clustering solution into two parts: the {\em definition} of clusters and the {\em algorithmic design} of efficiently clustering a given database according to the cluster definition.  Many existing work on clustering contribute to both fronts - e.g., $k$-means clustering defines clusters according to the distance between a point and the $k$ cluster centers, while density based clustering defines clusters according to how points are closely packed together.  Our objective in this paper is {\em not} to challenge or improve the cluster definitions of prior work, but rather to enable the second part - i.e., algorithmic design according to a given cluster definition - over an LBS with a limited $k$NN interface. From this perspective, our goal here is largely orthogonal to traditional clustering research.

\vspace{1mm}
{\noindent \bf Performance measure:}
There are two important performance measures for clustering over LBS: (1) the efficiency of clustering, and (2) the quality of clustering output $f(\cdot)$. For efficiency, the key bottleneck in our problem setting is the query-rate limit imposed by real-world LBS.  Thus, we focus on one efficiency measure in this paper: {\em query cost}, i.e., the number of queries the clustering algorithm has to issue in order to produce $f(\cdot)$.  Note that traditional efficiency measures, e.g., the computational and storage overhead of the clustering algorithm, are all secondary concerns in our problem because of the limited input size - note that the number of ``input'' tuples, i.e., tuples that can be retrieved from the underlying database, is inherently bounded by $k$ times the query cost, which is likely a small number in practice due to the query-rate limit.

In terms of clustering quality, we need to compare $f(D)$, i.e., the outputs of $f(\cdot)$ for all tuples in $D$, with a reference set of clustering IDs that can be either the ground truth or the output of a traditional clustering algorithm over the entire $D$.  In either case, the difference between $f(D)$ and the reference set can be measured in a variety of metrics commonly used in clustering research \cite{halkidi2001clustering}.  In this paper, we consider three metrics in experimental analysis: {\em Rand index}, {\em Jaccard index} and {\em Folkes and Mallows index} \cite{halkidi2001clustering}, respectively.

Let $P_1, P_2, \ldots P_h$ be the clusters produced by $f(D)$.  Assume $D = P_1 \cup \cdots \cup P_h$, as points deemed noise (i.e., with $f(t)$ being NULL) can be considered as all belonging to a ``noise'' cluster.  Let $C_1, C_2, \ldots C_{h^\prime}$ be the ground-truth clustering result, again with $D = C_1 \cup \cdots \cup C_{h^\prime}$.  The design of Rand, Jaccard, and Folkes and Mallows indices all consider the following four critical numbers:
\begin{itemize}
\item $a$, the COUNT of pairs of tuples in $D$, say $t, t^\prime$, that belong to same cluster according to both $P$ and $C$, i.e., $\exists i \in [1, h]$ and $j \in [1, h^\prime]$, such that $\{t, t^\prime\} \subseteq P_i$ and $\{t, t^\prime\} \subseteq C_j$.
\item $b$, the COUNT of pairs of tuples in $D$ that belong to different clusters according to both $P$ and $C$.
\item $c$, the COUNT of pairs of tuples in $D$ that belong to the same cluster according to $P$ but different ones according to $C$.
\item $d$, the COUNT of pairs of points that belong to different clusters according to $C$ but the same one in $P$.
\end{itemize}
The Rand Index measure ($R$), Jaccard Index ($J$) and Fowlkes and Mallows index ($FM$) are defined as below:

\begin{align}
    R = \frac{a + b}{a + b + c + d}, J = \frac{a}{a + c + d}, FM = \frac{a}{\sqrt{(a + c) \cdot (a + d)}} \nonumber
\end{align}

\subsection{Density Based Clustering}

Since most real-world LBS focus on 2D points, we consider density-based clustering, a popular class of techniques for low-dimensional data\cite{berkhin2006survey}.  Before discussing how to enable density-based clustering over LBS in the technical sections, here we briefly review its basic design in the traditional setting of a database with full access.

Density based clustering algorithms use the density property (e.g., reachability) of points to partition them into separate clusters. Specifically, the output depicts dense clusters (of points) separated by low-density regions. DBSCAN \cite{ester1996density} is the an example. It takes two parameters as input: $\epsilon$, the radius of a region under consideration, and $minPts$, the minimum number of points inside a region for it to be dense. Specifically, a point $t \in D$ is considered {\em core} if there are at least $minPts$ points within distance $\epsilon$ of $t$. Two points $t_0, t_r \in D$ are {\em reachable} from each other if there is a sequence of core points $t_1, \ldots, t_{r-1}$, such that $\forall i \in [0, r-1]$, $t_i$ and $t_{i+1}$ are within distance of $\epsilon$ from each other.

Given $\epsilon$ and $minPts$, the definition of a cluster becomes straightforward. Specifically, a core point $t$ is clustered together with all points in $D$ that are reachable from $t$.  If a point is not reachable from any other point, it becomes an outlier, i.e., noise.  Other density-based clustering techniques follow similar principles, but measure density in different ways, leading to different {\em definitions} of clusters.  For example, DENCLUE \cite{hinneburg1998efficient} defines the density of a point as a sum of the influence function of the other points in $D$. OPTICS\cite{berkhin2006survey} generalizes the density definition of DBSCAN by enabling different local densities with $\epsilon^\prime \leq \epsilon$\cite{berkhin2006survey}. 

As discussed earlier in this section, to enable clustering over LBS, we inevitably have to choose a cluster definition to follow.  For the purpose of this paper, we consider the simple $(\epsilon, minPts)$-based definition of DBSCAN, and aim to produce a cluster assignment function $f(\cdot)$ with $f(D)$ being as close to the clusters defined by DBSCAN as possible (as measured by the aforementioned metrics $R$, $J$ and $FM$).  One special note here is that, since $minPts$ is often set to be a small value such as $20$ in practice\cite{ester1996density}, we assume $minPts \leq k$ (as in the $k$NN interface offered by the LBS).  This way, whether $t$ is a core point can be determined by just one query (i.e., on $t$, by judging whether the $minPts$-th returned point is within $\epsilon$ from $t$).  In case $minPts > k$, one can always call the $k$NN based crawling algorithm \cite{YGZ16} to first crawl the $2\epsilon \times 2\epsilon$ square surrounding $t$ and then make the determination. 

\section{1D Case}
\label{sec:oneD}

We now consider how to enable clustering over the $k$NN interface of an LBS.  As discussed in the introduction, we start by developing HDBSCAN-1D for 1D data. The simplicity of the 1D setting allows us to highlight the basic idea of our sampling-based design. Additionally, the general technique we develop in the next section leverages the 1D algorithm as a subroutine for solving the high-dimensional problem. 

\subsection{1D Baseline and Problem}

Since we follow the cluster definition in DBSCAN, we start by considering a partitioning of the 1D space into  cells of equal width, $\epsilon$. We call such a cell ``dense'' if there are at least $minPts$ points in it, and ``sparse'' otherwise.  Note that, so long as one can somehow determine (by issuing $k$NN queries) which cells are dense and which are not, the result of DBSCAN can be almost faithfully replicated by joining adjacent dense cells to form a cluster. 

It is easy to determine the density of a single cell.  One simply needs to issue a $k$NN query $q$ at the center of the cell, and see whether the nearest $minPts$ points returned (recall from \S\ref{sec:datamodel} that $minPts \leq k$) all fall within the cell.  If so, the cell must be dense.  Otherwise it must be sparse.

Extending the solution to determine the density of all cells, however, is not easy.  A baseline solution here is to select cells uniformly at random, and issue queries at their centers to determine their density.  The problem, however, is its high query cost.  Note that any $k$NN query answer can ``cover'' at most $k/minPts$ dense cells.  Thus, generally speaking, if there are $D_\mathrm{c}$ core points in $D$, the number of queries required is at least $D_\mathrm{c} \cdot minPts/k$. Given the small $k$ offered by real-world LBS systems, this query cost tends to be prohibitively expensive in practice.

\subsection{Algorithm HDBSCAN-1D}
\noindent{\bf Key Idea:} To address the problem with the baseline solution, our main idea is to leverage the {\em locality} of cell densities - i.e., cells adjacent to each other are highly likely to have similar density values. This locality property, of course, is not new. It has been used in the design of many density-based clustering algorithms - e.g., DENCLUE \cite{hinneburg1998efficient} reduces the problem of density based clustering to kernel density estimations, essentially assuming neighboring cells to have similar densities according to a Gaussian mixture model.

Specifically, our HDBSCAN-1D differs from the baseline on how to deal with a query $q$ that returns $minPts$ points within a cell. In addition to marking the cell as dense, we also aim to (approximately) identify the {\em boundary} of the entire cluster surrounding the cell - i.e., the maximal sequence of cells, containing the one queried, that are all dense.  The rationale here, of course, is that according to the locality property, the number of such sequences is much smaller than the number of dense cells in the space.

Consider a query $q$ which returns at least {\em minPts} points within cell $q$ - note that here we use the same notation to represent the query and the cell without ambiguity, because a query we issue is always right at the center of the 1D cell.  Our idea is to first find a {\em sparse} cell to the left (resp. right) of $q$ in the 1D space, and then identify the dense cell immediate to the right (resp. left) of the sparse region as a  candidate for the left (resp. right) boundary of $q$'s cluster. We discuss this process in detail as follows.

First, we set an initial range $(a, b)$ based on the nearest sparse cells to $q$ that we already know. That is, both $a$ and $b$ are already discovered sparse cells with $a \leq q \leq b$.  If no sparse cell has been discovered, we can set $a$ and $b$ as the boundaries of the value domain. We start by finding the first dense cell to the right of $a$. To do so, we issue a query at $a + 1$. If $a + 1$ returns at least one point to its right, say at cell $a^{\prime}$, then we test the density of $a^{\prime}$ by issuing $a^{\prime}$ as a query. If $a^{\prime}$ is dense, we have accomplished our task. Otherwise, since we now know that $a^{\prime}$ is sparse, we can shrink the range to $(a^{\prime},b)$ and repeat this process.

Once we have identified the first dense cell to the right side of $a$, say $d$, our task now is to determine if $[d, q]$ consists solely of dense cells - i.e., if $d$ is the left boundary for the cluster containing $q$. We sample $c$ cells uniformly at random from $[d, q]$ to test if all of them are dense. If all $c$ cells turn out to be dense, we consider the range to be all-dense, i.e. continuous range of dense cells. Alternatively, we find a sparse cell $d^{\prime} \in  [d, q]$ - as soon as we discover $d^{\prime}$, we repeat the entire process with an updated initial range of $(d^{\prime}, b)$.

One can see that this process eventually leads to a range $[d, e]$ containing $q$ that pass the $c$-sample test of being consecutive dense cells.  What we do next is to sample uniformly at random a cell for which we cannot yet determine density, and query the cell to repeat the above-described process. Once again, this can be repeated until the density nature of all cells have been determined, or until all query budget has been exhausted.

\begin{algorithm}[!htb]                                                                             
\caption{HDBSCAN-1D}                                                                                
\begin{algorithmic}[1]                                                                              
\label{alg:hdbscan1D}                                                                               
\STATE {\bf while} query budget is not exhausted 
  \STATE \hindent Issue $k$NN query over randomly chosen unvisited cell $q$ 
  \STATE \hindent {\bf if} $q$ is dense                                                             
    \STATE \hindent[2] (a, b) = range containing $q$ where both $a$ and $b$ are already discovered sparse cells.
    \STATE \hindent[2] $l$, $r$ = left and right boundary discovered using binary search inside range [a+1, q] and [q, b-1] respectively.
    \STATE \hindent[2] Add the cell range [$l$, $r$] to dense segment list 
\STATE Output the dense segments (clusters) identified so far.                                      
\end{algorithmic}                                                                                   
\end{algorithm}     

\vspace{1mm}
\noindent {\bf Query Cost Analysis:} We start by considering an ideal case where all tuples in the database belong to one of the $h$ clusters - i.e., there are no outlier points.  During the initial search of the first dense cell to the right of $a$, there are only two possible outcomes for each query we issue: either it shows the cell to be dense and triggers a $c$-cell density test, or it returns no tuple to its right side at all. 

For the first case (i.e. the $c$-cell density test), note that each failed test identifies a new cluster - i.e., we conduct the test at most $h$ times, consuming $O(h \cdot c)$ queries in the worst-case scenario. When the query returns no point to its right, however, the situation can be more complex. Note that if a query $a^{\prime}$ returns no point to its right, then all the $k$ nearest neighbors to $a + 1$ are on its left side. What this triggers is a process that we refer to as {\em exponential search}. Specifically, let $\ell$ be the maximum distance between $a^{\prime} + 1$ and the $k$ returned points. Since we are now certain that no tuple resides within $[a^{\prime}, a^{\prime} + \ell - 1]$, the search space is shrunk to $[a^{\prime} + \ell, b]$ - i.e., the next query we issue will be at $a^{\prime} + \ell$. This query either returns a point to the right side of $a$, or proves that no point resides in $(a, a + 3\ell)$. If it returns a point to the right side of $a$, the $c$-sample test is triggered, with query cost falling within the $O(h \cdot c)$ queries in the above analysis. Otherwise, if $a + \ell$ still does not reveal any point to the right of $a$, i.e., our next queries to issue would be $a + 3\ell, a + 7\ell, a + 15\ell, \ldots$ - representing exponential growth of the query value. Note that each exponential search consumes $O(\log N/\epsilon)$ queries. The exponential search process can occur at most $h + 1$ times, corresponding to the $h + 1$ empty segments in between the $h$ clusters. As such, the overall query cost becomes $O(h \cdot c + h \cdot O(\log N/\epsilon))$.

\vspace{1mm}
\noindent {\bf Handling Noisy Points:} While the technique described above shows a significantly reduced query cost according to the above analysis, it actually has an important problem masked by an assumption made in the analysis - i.e., there is no outlier in $D$ and every point belong to a cluster.  Note that if $a^{\prime}$ returns a point to its right which nonetheless does not turn out to be a dense cell, then the next query issued would be $a^{\prime} + 1$ (instead of much further to the right as in the case of exponential search). In other words, in the worst case scenario where every cell is filled with fewer than {\em minPts} but at least one point, the technique might have to enumerate all cells between $a$ and the left boundary of the cluster containing $q$.

To address the problem, we introduce a binary search process for this scenario. Specifically, we start by querying $(a^{\prime}+ q)/2$. If it is dense, we move to the left (i.e., $(3 a^{\prime} + q)/4$, $(7 a^{\prime} + q)/8$, etc., in order). If it is sparse, we move to the right. One can see that this process always terminates when we discover a dense cell that has its immediate neighbor to the left being sparse. In other words, we have discovered the left boundary of a cluster - whether the cluster is the one surrounding $q$ will be verified by the $c$-sample test.

One can see that each execution of this binary search process returns one of the two boundaries for a cluster. Thus, it is executed for $O(h)$ times in the worst case. In other words, the overall query cost remains $O(h \cdot c + h \cdot O(\log N/\epsilon))$.



\begin{figure*}[!t]
  \begin{minipage}[t]{0.23\linewidth}
    \centering
    \includegraphics[scale=0.4]{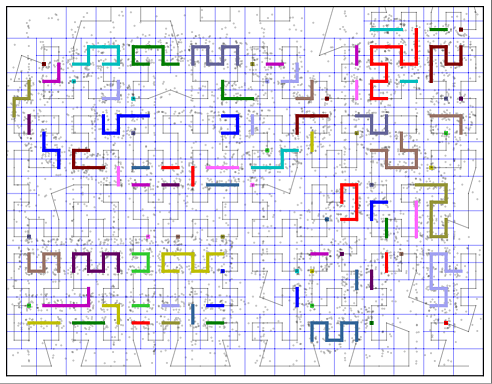}
    \caption{Mapping 2D space to 1D using Adaptive SFC}
    \label{fig:ch7_10k_step1}
  \end{minipage}
  \hspace{1mm}
  \begin{minipage}[t]{0.23\linewidth}
    \centering
    \includegraphics[scale=0.4]{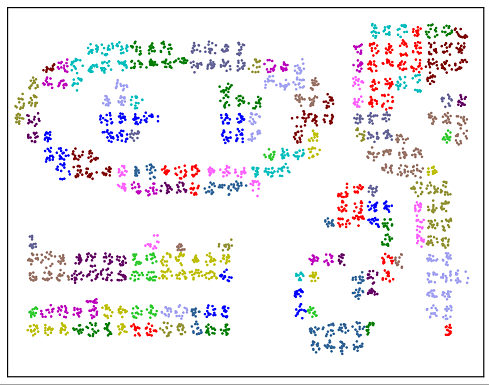}
    \caption{After clustering in 1D space}
    \label{fig:ch7_10k_step2}
  \end{minipage}
  \hspace{1mm}
  \begin{minipage}[t]{0.23\linewidth}
    \centering
    \includegraphics[scale=0.2]{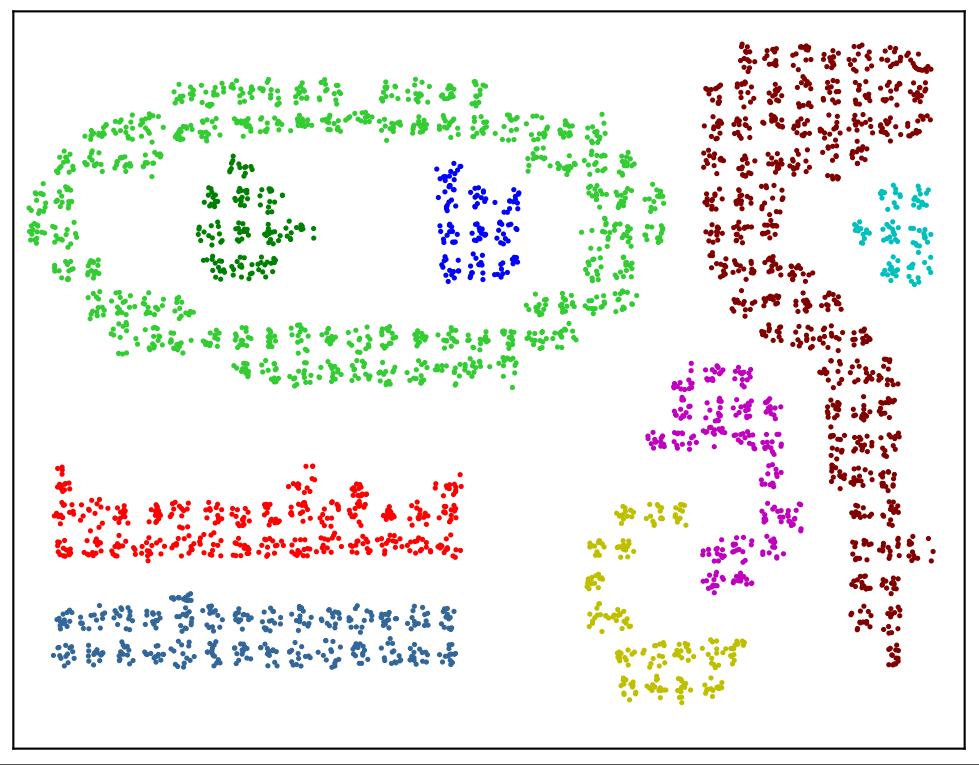}
    \caption{Merging nearby clusters in 2D space}
    \label{fig:ch7_10k_step3}
  \end{minipage}
  \hspace{1mm}
  \begin{minipage}[t]{0.23\linewidth}
    \centering
    \includegraphics[scale=0.4]{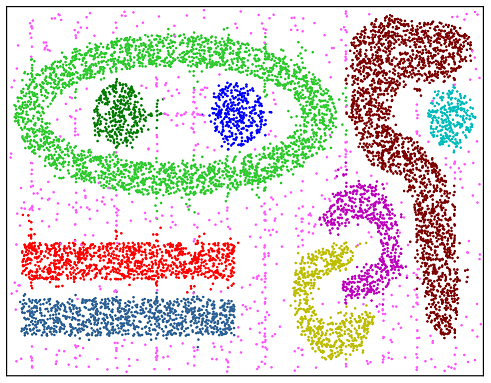}
    \caption{Assigning new points to closest cluster}
    \label{fig:ch7_10k_step4}
  \end{minipage}
\end{figure*}

\section{HDBSCAN} \label{sec:generalCase}

In this section, we consider the general case of clustering higher-dimensional data through a restrictive $k$NN interface.  While it will be clear from the discussions that our results can be directly extended to data of any dimension, the focus of this paper is on 2D data because of its prevalence among real-world LBS.  

\subsection{Overview}

One can see from the design of HDBSCAN-1D that the ``boundary'' pursuit idea used there cannot be easily extended to higher dimensions, because instead of having just 2 bounds (left and right) as in 1D, there may be a large number of cells bounding a 2D (or higher-D) cluster of an arbitrary shape, making the exact discovery of them extremely expensive in query cost.

\begin{figure}[!h]
  \centering
  \includegraphics[scale=0.95]{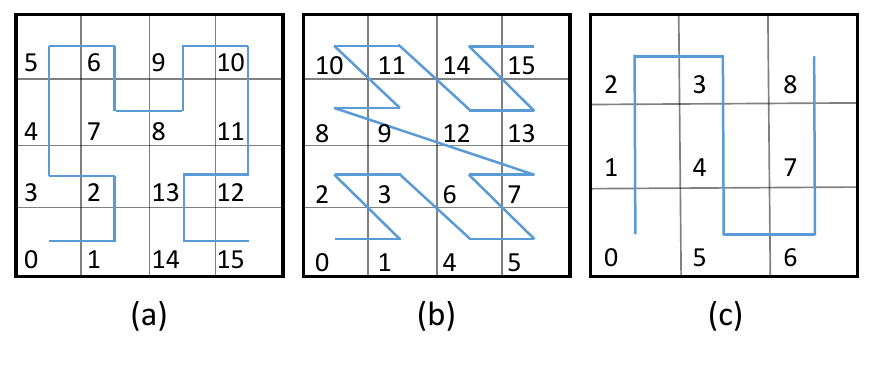}
  \caption{Illustration of popular SFC: (a) Hilbert curve (b) Z-curve (c) Peano curve}
  \label{fig:popularSFC}
\end{figure}

\begin{figure*}[!t]
  \begin{minipage}[t]{0.48\linewidth}
      \centering
      \includegraphics[scale=1.0]{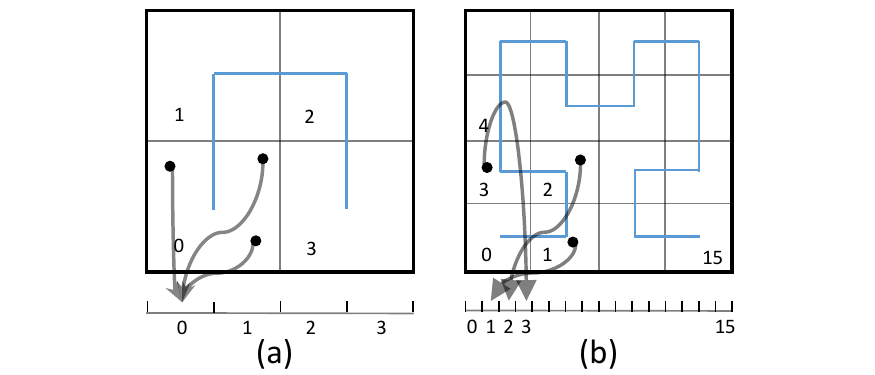}
      \caption{Mapping 2D points to 1D through Hilbert SFC}
      \label{fig:hilbertCurveMapping}
  \end{minipage}
  \hspace{1mm}
  \begin{minipage}[t]{0.48\linewidth}
      \centering
      \includegraphics[scale=1.0]{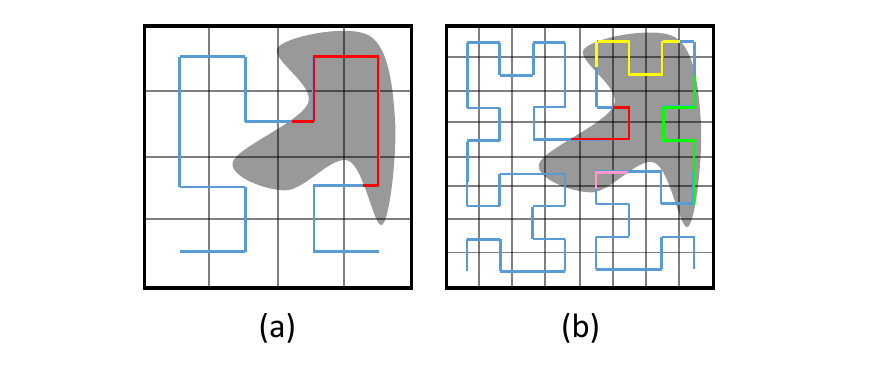}
      \caption{Impact of Grid Size on Clusters}
      \label{fig:impactOfGridSize}
  \end{minipage}
\end{figure*}

To address this challenge, our idea is to first map 2D data into a 1D space, and then call upon HDBSCAN-1D to perform clustering as described in \S\ref{sec:oneD}.  To enable the mapping, we use a specially designed {\em Space Filling Curve} (SFC) detailed later in the section.  The general concept of SFC\cite{xu2014optimality,mokbel2003analysis,moon2001analysis}  is illustrated in Figure~\ref{fig:popularSFC}.  More specifically, for each 2D point $t$, we records as its mapped 1D coordinate $S(t)$ the point on the SFC that is closest in distance to $t$ (in the 2D space).  With this design, each query HDBSCAN-1D decides to issue is mapped to its corresponding 2D coordinates and issued to the underlying 2D LBS, while every returned 2D tuple $t$ is mapped back to its 1D coordinate $S(t)$ for HDBSCAN-1D to process.   One can see that this two-way mapping enables the seamless execution of HDBSCAN-1D over the 2D space.  In addition, it ensures that two tuples $t_1, t_2 \in D$ (indeed, their mappings $S(t_1)$ and $S(t_2)$) belonging to the same cluster produced by HDBSCAN-1D should also be clustered together in the 2D space. Figure~\ref{fig:hilbertCurveMapping} shows a mapping from 2D to 1D space using Hilbert curve.

Nonetheless, this SFC-based mapping also introduces a major challenge to the clustering design:  Since no SFC can guarantee that two points close together in 2D are always close in 1D \cite{xu2014optimality}, we are left with the possibility that one 2D cluster may be partitioned into multiple 1D clusters after the mapping.  To address the problem, we introduce an additional step of {\em merging} the ``mini-clusters'' produced by HDBSCAN-1D.  Specifically, recall from \S\ref{sec:oneD} that HDBSCAN-1D outputs $h$ mini-clusters as ranges $C_i: [a_i, b_i]$ ($i \in [1, h]$).  We consider for each pair\footnote{While this pairwise-testing appears to be an expensive (quadratic) process, note that the cost incurred here is local computational overhead instead of query cost, the bottleneck in our problem setting.} of mini-clusters whether they should be merged (with details discussed later in the section), according to the 2D points we have observed in each mini-cluster.  This merging process essentially produces a many-to-one mapping $C_i \to C^\prime_j$ ($i \in [1, h]$, $j \in [1, h^\prime]$), so each 1D range $C_i$ is labeled with a final cluster ID from $1$ to $h^\prime$ ($h^\prime \leq h$).

In the following four subsections, we shall discuss, respectively, the three critical steps for HDBSCAN: (1) the design of SFC for calling HDBSCAN-1D, (2) the merging of mini-clusters, and (3) the generation of clustering function $f(\cdot)$. 
Figures~\ref{fig:ch7_10k_step1} -~\ref{fig:ch7_10k_step4} shows the output of each steps described above for Chameleon t-7.10k dataset. Points that are detected as noise are colored in pink.

\subsection{Mapping 2D to 1D}

\noindent{\bf Baseline Solution and Problem:} To design the mapping from 2D to 1D, a baseline solution is to first partition the 2D space into $\epsilon \times \epsilon$ grid cells, and then apply standard SFCs such as Hilbert curve, Peano Curve, Z-curve, etc.  This way, the mapped 1D space is naturally partitioned to cells of width $\epsilon$, exactly matching the need of HDBSCAN-1D.  Meanwhile, the mini-clusters produced by the algorithm, when mapped back to 2D, will consist of adjacent $\epsilon \times \epsilon$ cells, closely approximating the $\epsilon$-radius neighborhood considered by the original DBSCAN.

A problem with this baseline solution, however, is the large number of grid cells defined by the mapping.  Recall from the query-cost analysis in \S\ref{sec:oneD} that the number of queries required by HDBSCAN-1D is proportional to $h$, the number of clusters - in this case mini-clusters - identified by the algorithm.  From the illustration in Figure~\ref{fig:impactOfGridSize}, one can clearly see that the more fine-grained the 2D grid cells are, the more mini-clusters it will partition a true 2D cluster into.  When the granularity is down to $\epsilon \times \epsilon$ as in the baseline, the number of mini-clusters produced may far exceed the number of real clusters (i.e., $h \gg h^\prime$), leading to a very large query cost.

On the other hand, it is also important to note that we cannot arbitrarily enlarge the grid cell size to reduce the count.  To understand why, note that the mapping from 2D to 1D might project two points at the two opposite ends of a 2D cell to the same 1D coordinate. For example, consider the bottom left cell in Figure~\ref{fig:hilbertCurveMapping}. Both the top left corner and the bottom right corner of the cell will be mapped to the 1D point at the center bottom of the 2D cell.  This is usually not a problem for clustering when each cell is small (e.g., $\epsilon \times \epsilon$). However, if we make the cell size too large, then this mapping-induced error might cluster together two points far from each other, adversely impacting the quality of clustering results. 

\begin{figure}[!t]
  \centering
  \includegraphics[scale=1.5]{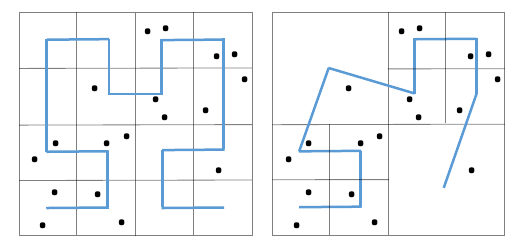}
  \caption{Standard vs Adaptive SFC}
  \label{fig:sfcFixedVsAdaptive}
\end{figure}

\vspace{1mm}
\noindent{\bf Adaptive Space Filling Curve (SFC):} To address the problem, our key idea here is to introduce the concept of an {\em adaptive SFC}, which combine larger grid cell sizes on sparser regions, in order to reduce the overall query cost, with smaller grid sizes on denser regions, in order to gain enough resolution to separate clusters from each other and outlier points. Figure~\ref{fig:sfcFixedVsAdaptive} depicts an example of such a space filling curve.

Note that the shape of the adaptive SFC depends on the underlying data distribution.  Since we do not have prior knowledge of the data distribution, we can no longer pre-define this SFC and its corresponding 2D to 1D mapping.  Instead, we have to construct the adaptive SFC on-the-fly and adjust the mapping as the SFC changes.  Specifically, this online process can be described as follows.
\begin{itemize}
\item We start with the largest possible grid cells, i.e., by partitioning the entire space into four cells, as shown on the root node in Figure~\ref{fig:sfcFixedVsAdaptive}.  Based on this initial SFC, we start the execution of HDBSCAN-1D.  Note that the design of the adaptive SFC is transparent to HDBSCAN-1D - it simply takes a 1D space as input and is oblivious to how large the cell sizes are in the 2D space.
\item For every query $q$ issued by HDBSCAN-1D, we identify the 2D cell $q$ falls into, say a cell of size $c \times c$, and issue a query $q^\prime$ at the center of the cell.  If the $minPts$-th ranked point $q^\prime$ returns is farther than $\sqrt{2}c/2$ from the cell center, it means that there can be no more than $minPts$ points within the $c \times c$ cell, and we do not need to partition it.  Otherwise, we partition the cell into four cells, as demonstrated in Figure~\ref{fig:sfcFixedVsAdaptive}.
\item Note that once we decide to further partition a cell, the adaptive SFC changes, and so is the 2D to 1D mapping (as it ``inserts'' a number of 1D cells to the domain).  Thus, starting from the next step in the execution of HDBSCAN-1D, we follow the new mapping for translating the 1D query and the 2D query answers.
\end{itemize}

%

\begin{figure}[!t]
  \centering
  \includegraphics[scale=0.9]{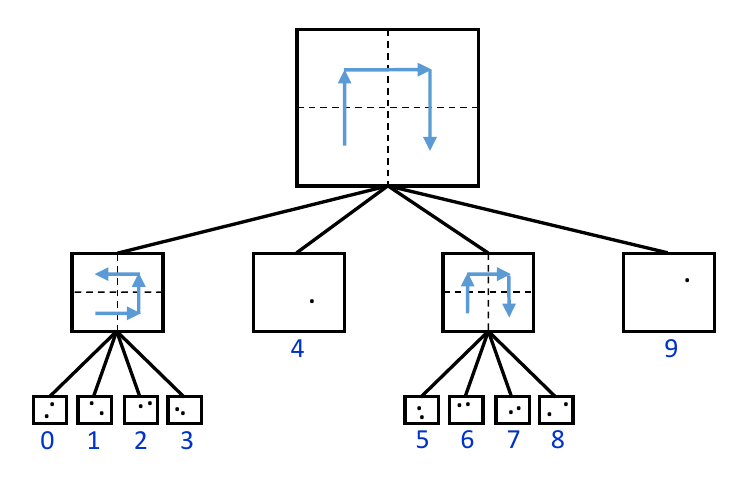}
  \caption{Adaptive SFC} 
  \label{fig:adaptiveSFCQuadTree}
\end{figure}


\subsection{Merging Mini-Clusters}
\label{sec:generalCase_MergeNeighborClusters}

We now consider how to merge the mini-clusters generated by HDBSCAN-1D to real 2D clusters.  A simple approach here is to merge mini-clusters containing grids that neighbor each other in 2D (i.e. grids that share an edge). A problem with this solution, as we found through experiments, is that it is likely to merge two clusters into one if the clusters happen to be close to each other. Setting the $\epsilon$ value very small might solve this, but this will also increase the query cost. To overcome this problem, we select a subset of points from each clusters as representative to compute the inter cluster distance. This is similar in nature to the concept of using fixed set of representative points to measure cluster distance in CURE \cite{berkhin2006survey}, a hierarchical clustering algorithm.  Specifically, we compute the $l$-distance between two mini-clusters - i.e., we identify all points in the two mini-clusters that have been observed in previous query answers.  Then, we select the top-$l$ pairs of points with the minimum distance from each other, and compute their average distance.  We merge the two mini-clusters if their $l$-distance falls below a pre-determined threshold. Setting a small value of $l$ would split larger clusters as they do not capture the shape of the clusters. Empirically, we found that setting $l$ to {\em minPts/2} provided best results. 


\begin{algorithm}[!htb]
\caption{HDBSCAN-2D}
\begin{algorithmic}[1]
\label{alg:hdbscan2D}
\STATE {\bf Input:} Tree $t$
\STATE {\bf while} query budget is not exhausted 
  \STATE \hindent Issue $k$NN query over unvisited node $n$ chosen randomly 
  \STATE \hindent {\bf if} $n$ is dense
 \STATE \hindent[2] Find dense segment containing $n$ using Algorithm~\ref{alg:hdbscan1D} such that none of $t$'s leaf nodes get partitioned in that process.
 \STATE \hindent[2] Add the dense segment to mini-cluster list.
\STATE Merge min-clusters using top-$l$ distance 
\STATE Output the dense clusters identified so far
\end{algorithmic}
\end{algorithm}

\subsection{Clustering of New Points}
Recall that our objective is to develop a clustering function $f(\cdot)$ that emulates the output of density-based clustering algorithms such as DBSCAN. Given a new point $f(\cdot)$ can then be used to identify its cluster affiliation. For each new point, we first map it to 1D space using the adaptive SFC and check whether it belongs to any of the ``mini-clusters'' previously discovered. The point is then assigned to the final cluster generated by the merging procedure from \S\ref{sec:generalCase_MergeNeighborClusters}. 

\section{Experimental Results}
\label{sec:exp}

\subsection{Experimental Setup}

\begin{figure*}[!ht]
  \begin{minipage}[t]{0.23\linewidth}
    \centering
    \includegraphics[scale=0.4]{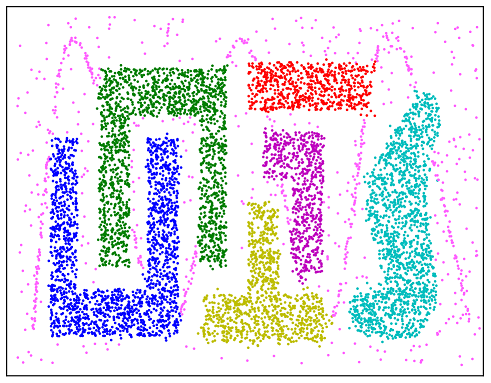}
    \caption{Chameleon t4.8k Dataset}
    \label{fig:chamaleon_4_8k}
  \end{minipage}
  \hspace{1mm}
  \begin{minipage}[t]{0.23\linewidth}
    \centering
    \includegraphics[scale=0.26]{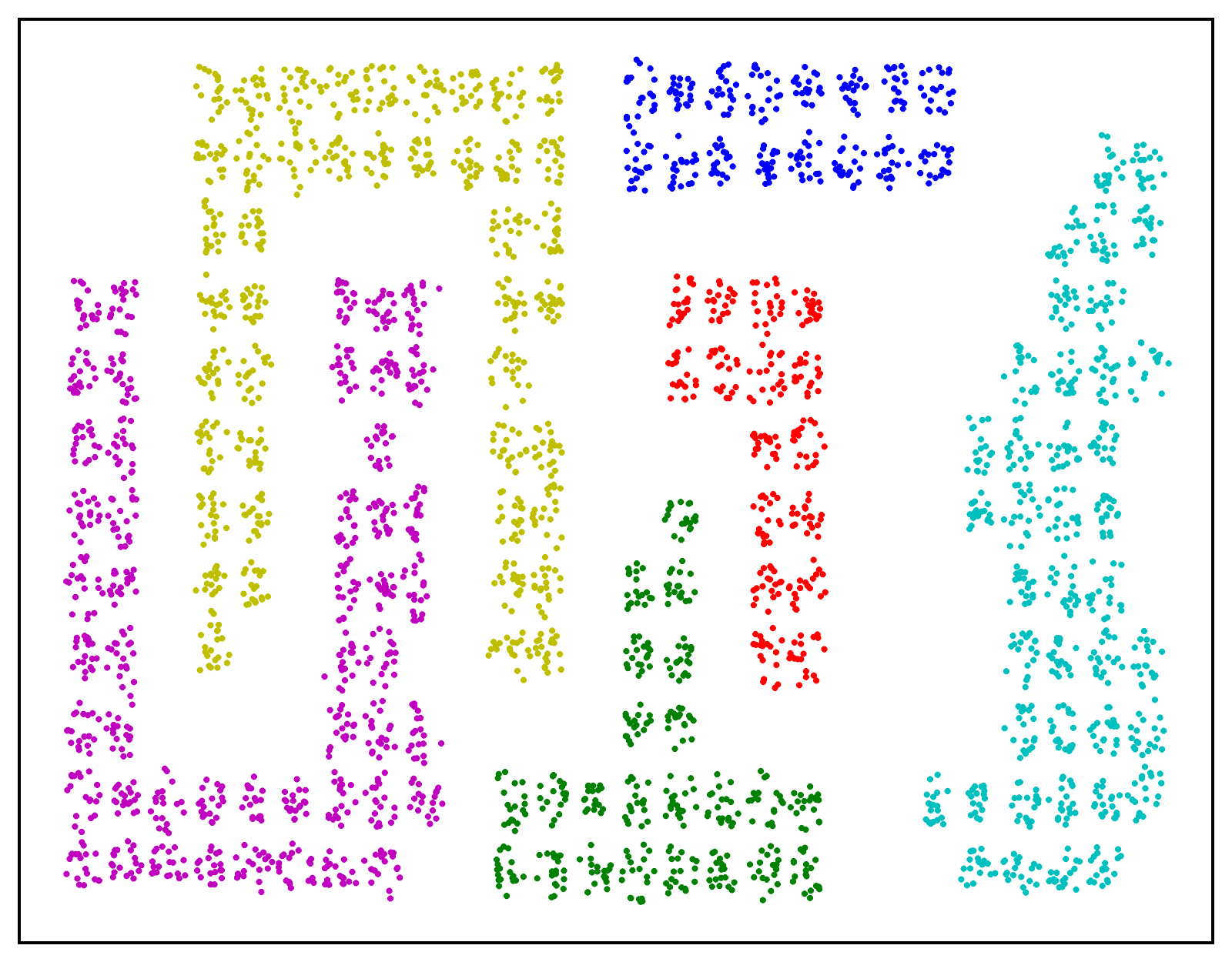}
    \caption{Output of HDBSCAN}
    \label{fig:chamaleon_4_8k_hdbscan}
  \end{minipage}
  \hspace{1mm}
  \begin{minipage}[t]{0.23\linewidth}
    \centering
    \includegraphics[scale=0.4]{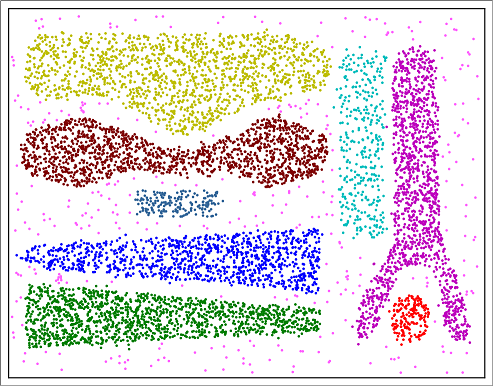}
    \caption{Chameleon t8.8k Dataset}
    \label{fig:chamaleon_8_8k}
  \end{minipage}
  \hspace{1mm}
  \begin{minipage}[t]{0.23\linewidth}
    \centering
    \includegraphics[scale=0.4]{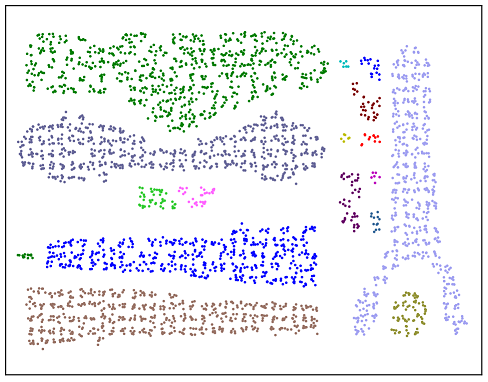}
    \caption{Output of HDBSCAN}
    \label{fig:chamaleon_8_8k_hdbscan}
  \end{minipage}
\end{figure*}

\begin{figure*}[!ht]
  \begin{minipage}[t]{0.23\linewidth}
    \centering
    \includegraphics[scale=0.19]{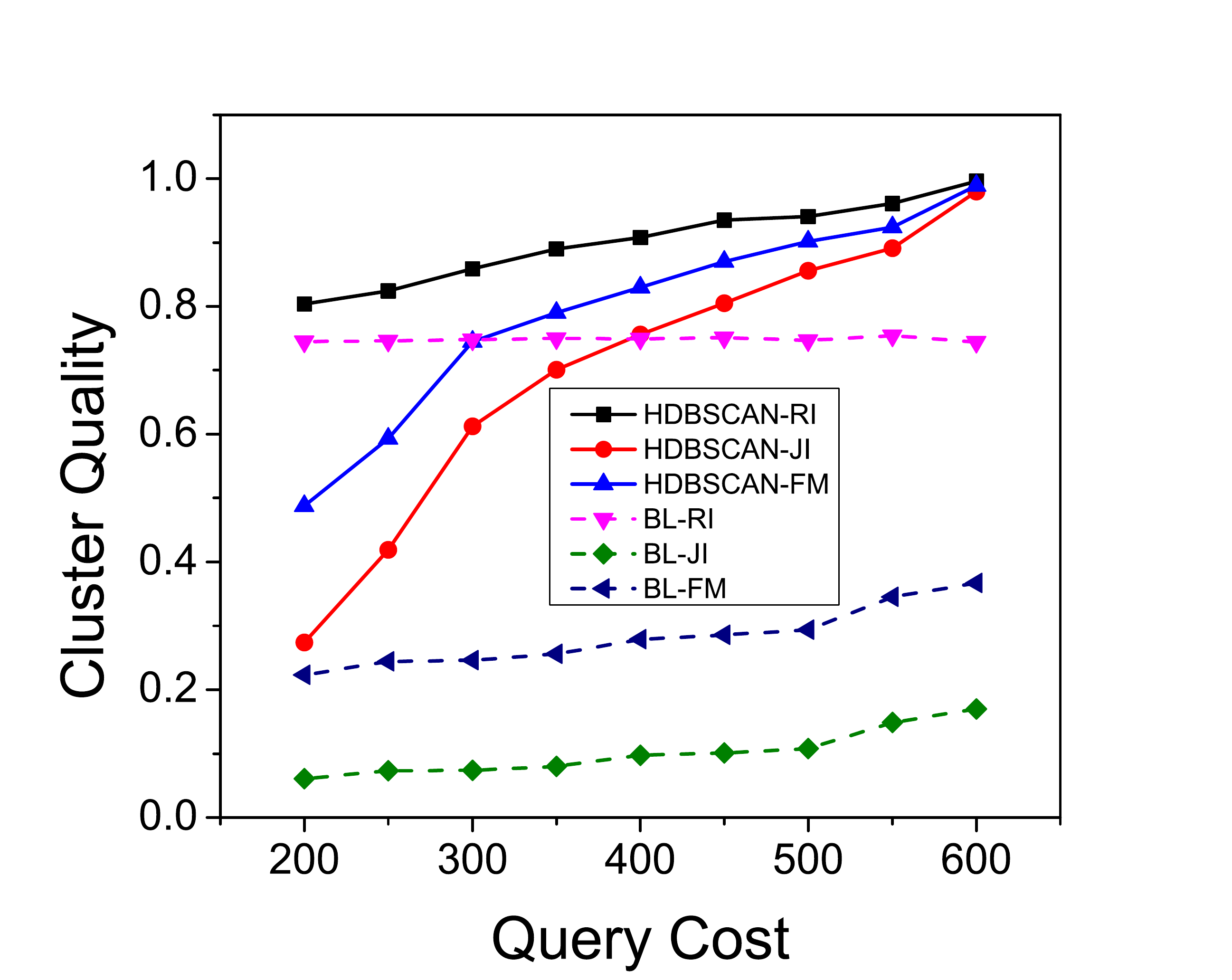}
    \caption{Clustering generated by baseline Algorithm}
    \label{fig:ch_7_10_baseline}
  \end{minipage}
  \hspace{1mm}
  \begin{minipage}[t]{0.25\linewidth}
    \centering
    \includegraphics[scale=0.2]{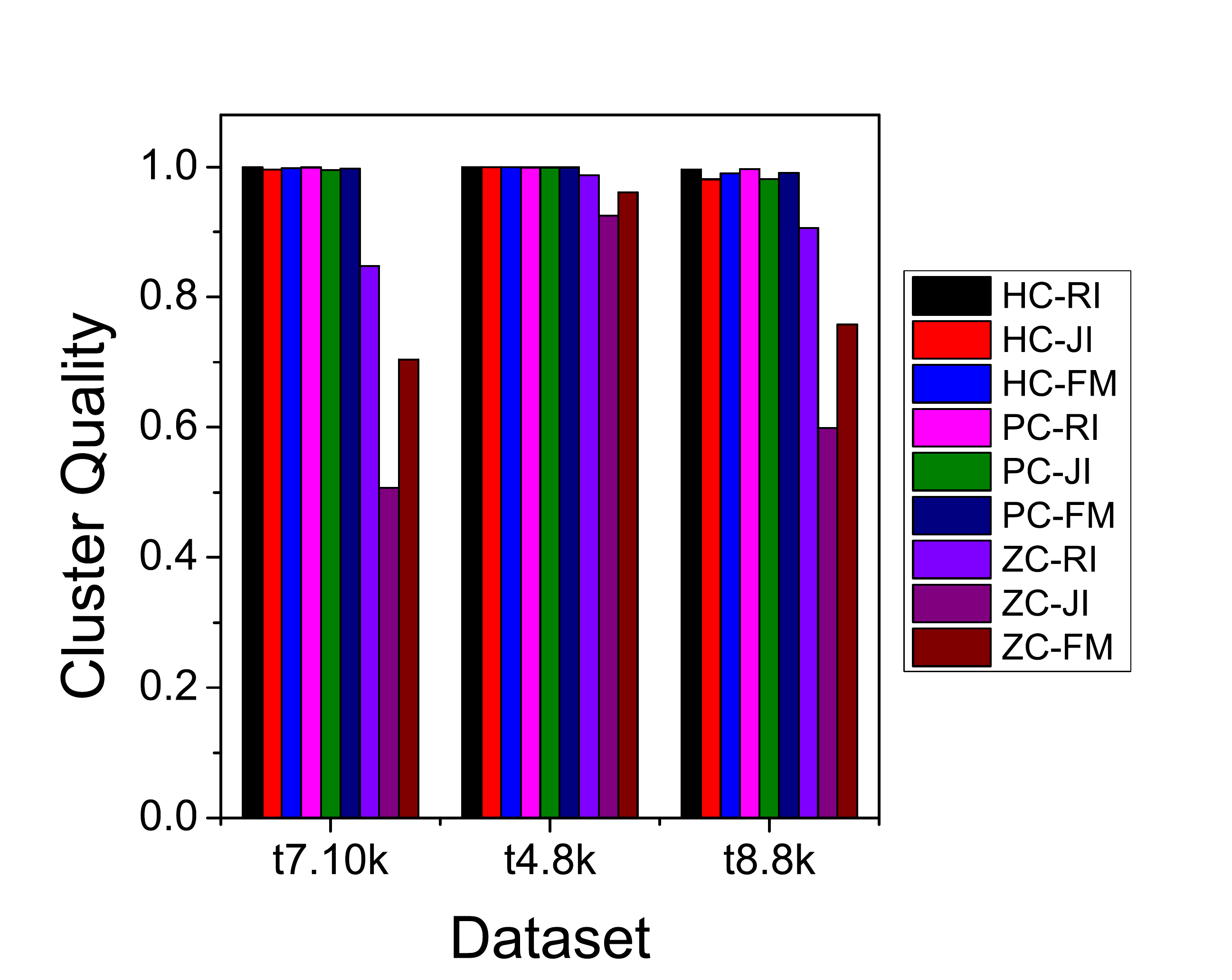}
    \caption{Cluster Quality for different SFCs}
    \label{fig:clusterQualityVsSFC}
  \end{minipage}
  \hspace{1mm}
  \begin{minipage}[t]{0.20\linewidth}
    \centering
    \includegraphics[scale=0.2]{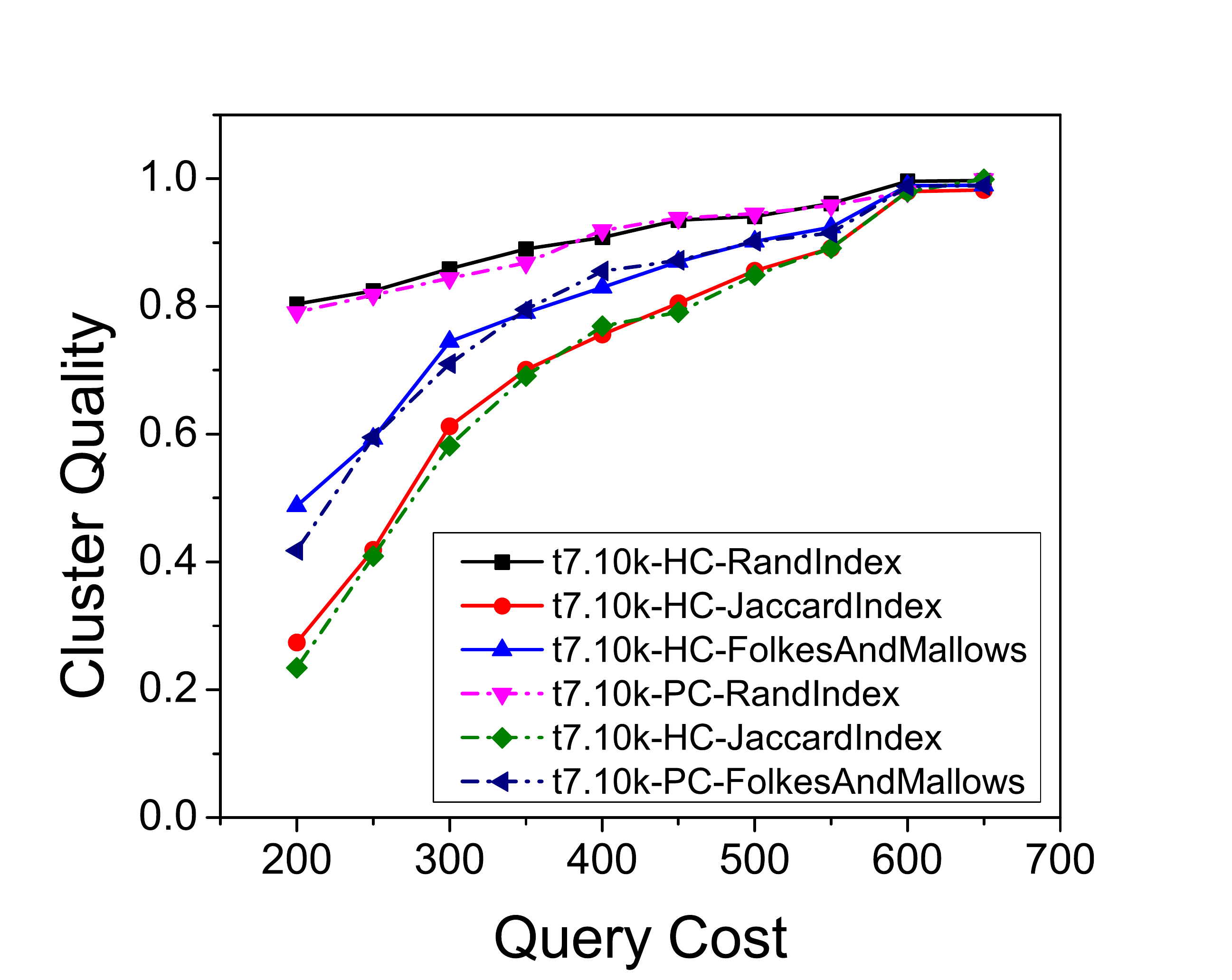}
    \caption{Cluster Quality Vs Query Cost}
    \label{fig:ch_clusterQualityVsQueryCost}
  \end{minipage}
  \hspace{8mm}
  \begin{minipage}[t]{0.25\linewidth}
    \centering
    \includegraphics[scale=0.2]{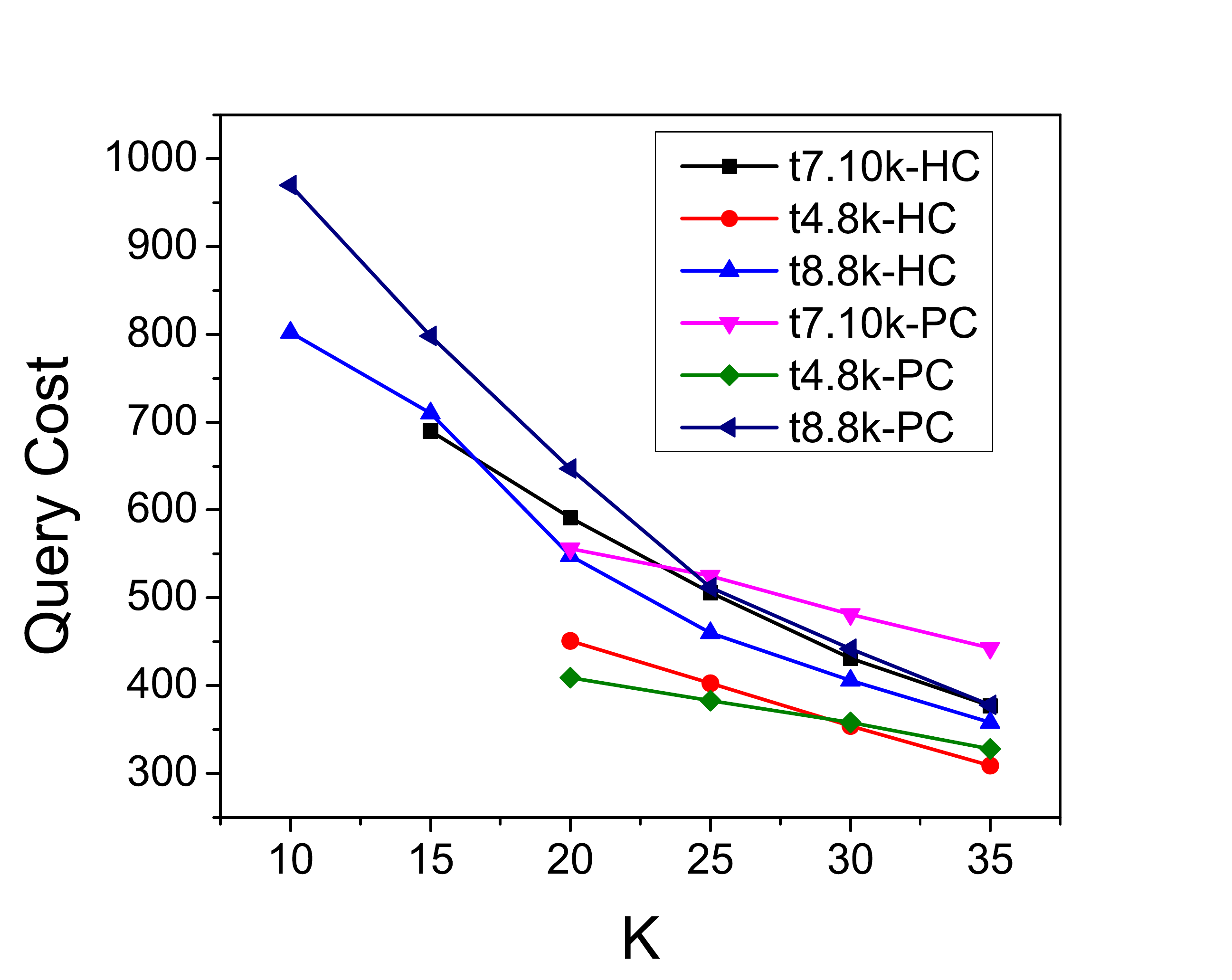}
    \caption{Varying K}
    \label{fig:ch_queryCostVsK}
  \end{minipage}
\end{figure*}

\begin{figure*}[!ht]
  \begin{minipage}[t]{0.23\linewidth}
    \centering
    \includegraphics[scale=0.2]{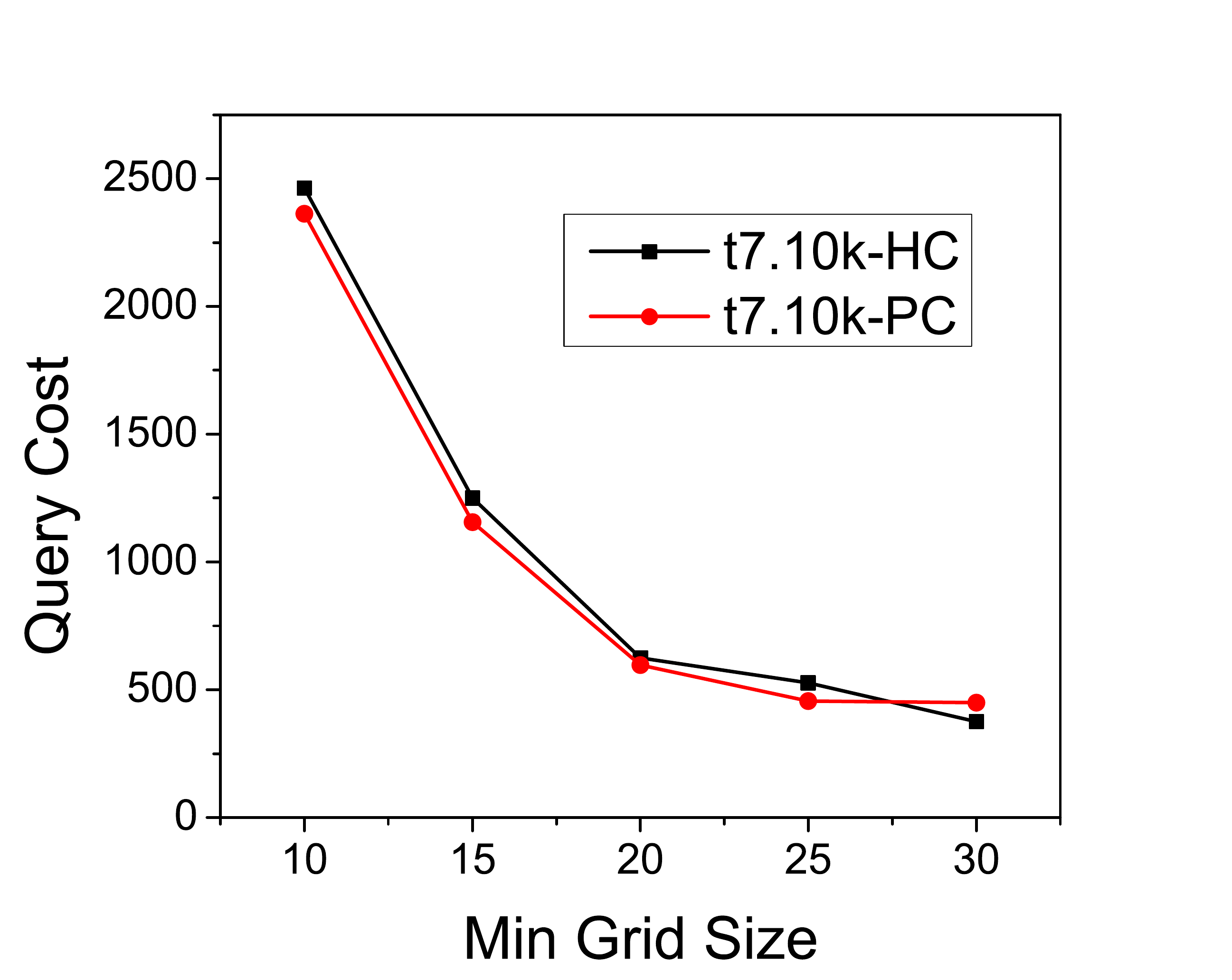}
    \caption{Varying min. grid size}
    \label{fig:ch_queryCostVsMinGridSize}
  \end{minipage}
  \hspace{1mm}
  \begin{minipage}[t]{0.23\linewidth}
    \centering
    \includegraphics[scale=0.2]{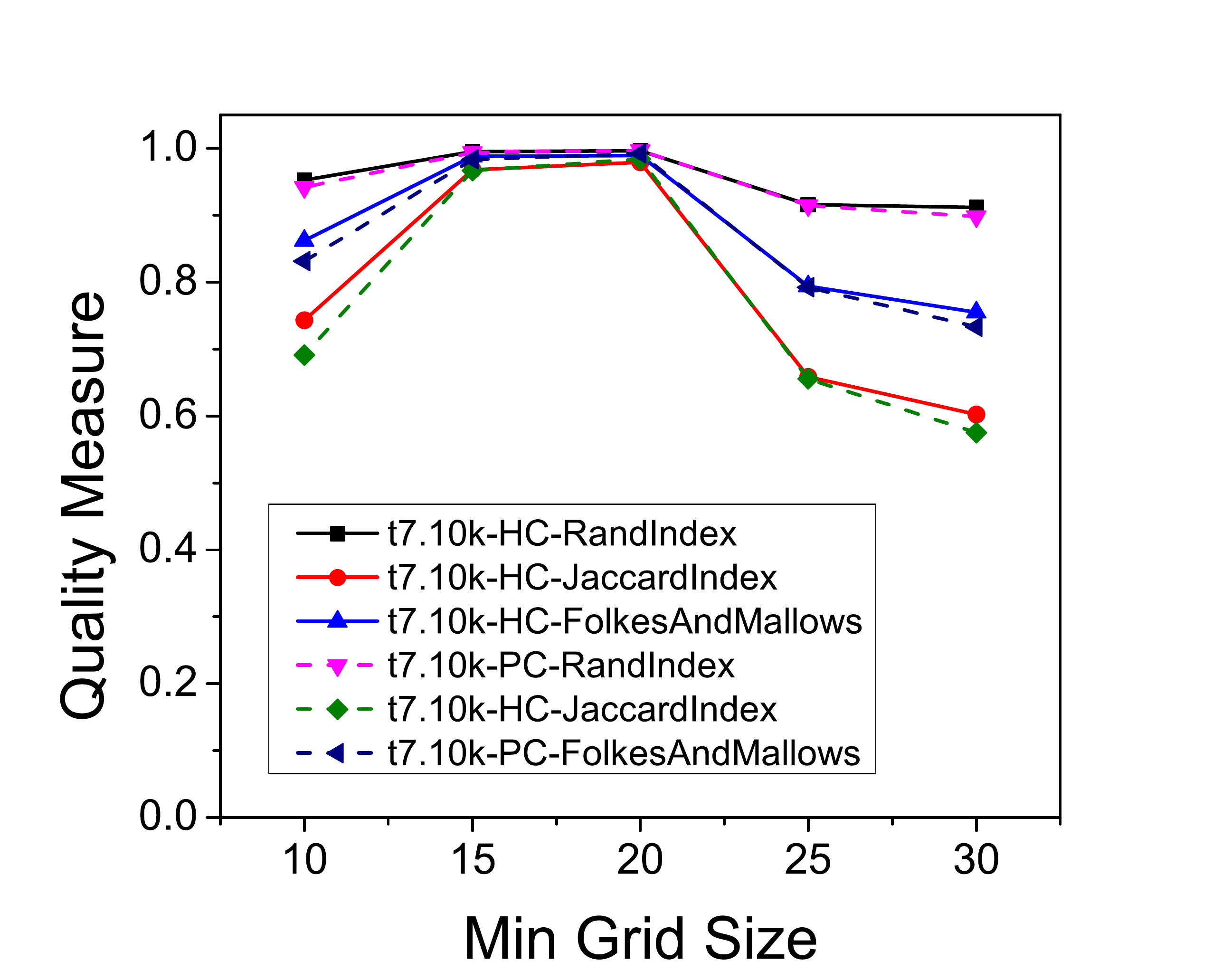}
    \caption{Varying min. grid size}
    \label{fig:ch_clusterQualityVsMinGridSize}
  \end{minipage}
  \hspace{3mm}
  \begin{minipage}[t]{0.26\linewidth}
    \centering
    \includegraphics[scale=0.2]{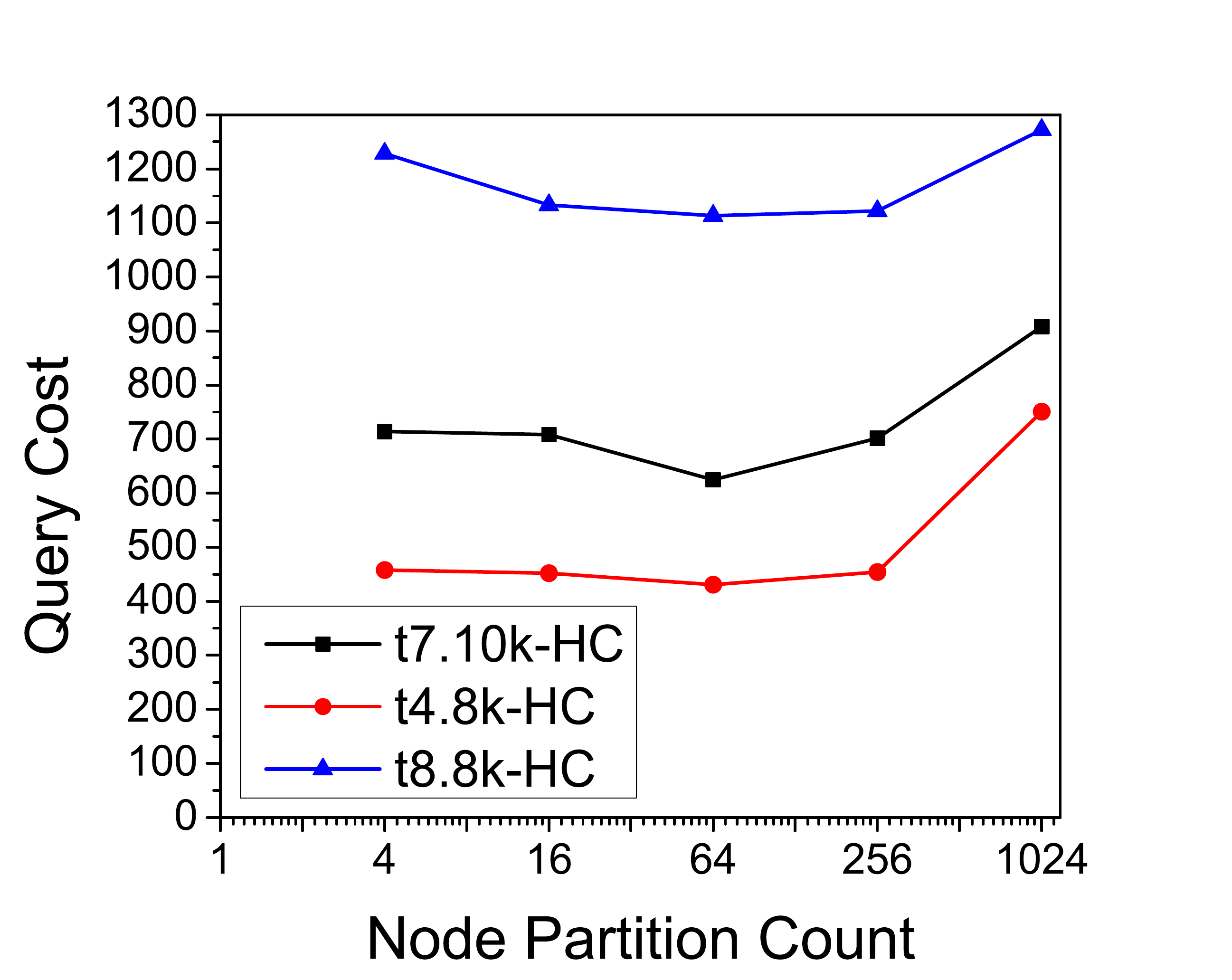}
    \caption{Varying number of Partitions per node} 
    \label{fig:ch_queryCostVsNodePartitionCount}
  \end{minipage}
  \hspace{1mm}
  \begin{minipage}[t]{0.23\linewidth}
    \centering
    \includegraphics[scale=0.2]{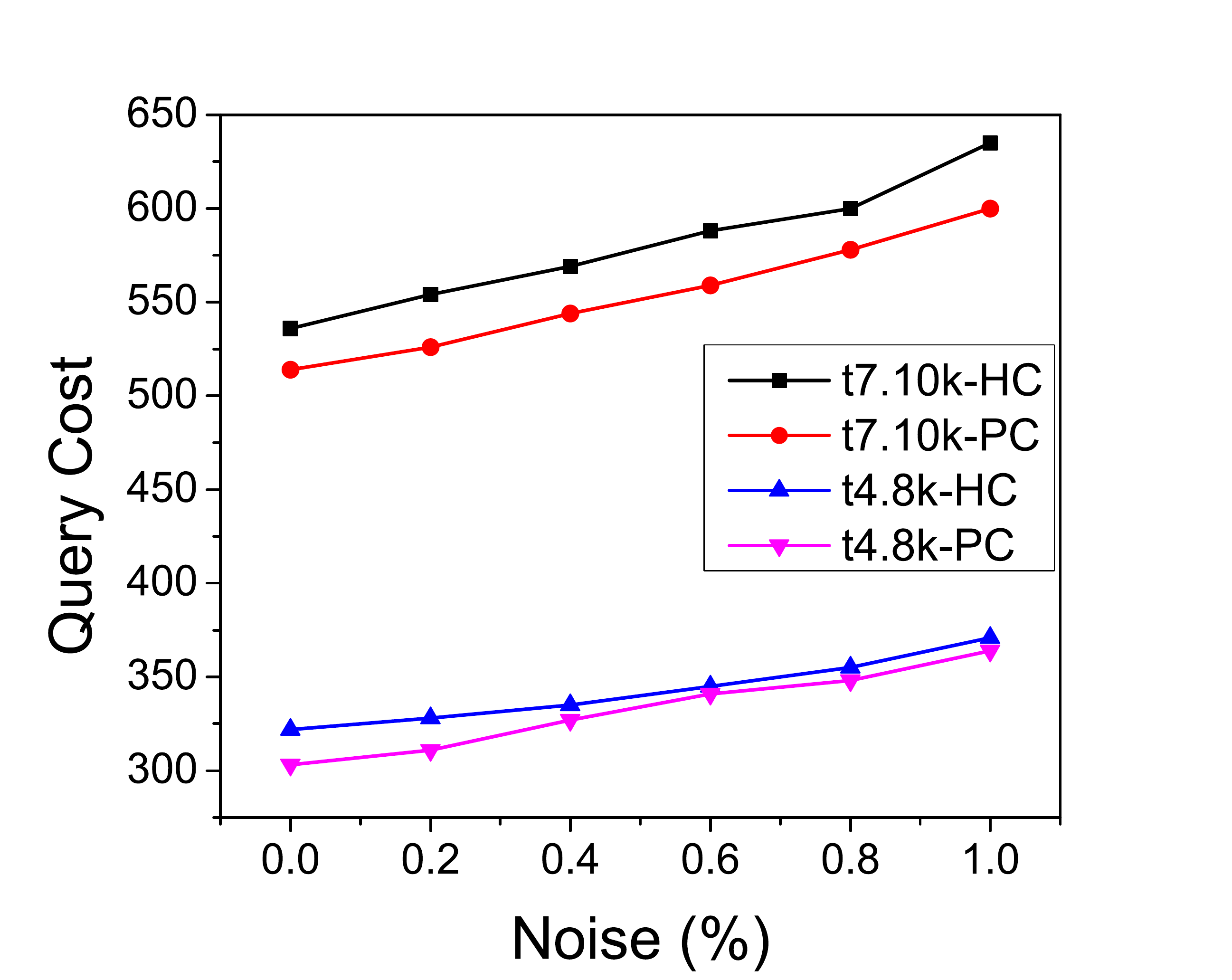}
    \caption{Varying percentage of Noise in dataset}
    \label{fig:ch_queryCostVsNoise}
  \end{minipage}
\end{figure*}

\noindent{\bf Hardware and Platform:}
All our experiments were performed on a quad-core 2.5 GHz Intel i7 machine running Ubuntu 14.10 with 16 GB of RAM. The algorithms were implemented in Java.

\noindent{\bf Benchmark Datasets:}
We evaluated the clustering quality generated by HDBSCAN using the widely used Chameleon benchmark datasets \cite{karypis1999chameleon}. They contain 2D points along with clusters of different shapes, densities, sizes and noises. We tested HDBSCAN on three Chameleon datasets: Chameleon-t7.10k, t4.8k and t8.8k. Note that for offline experiments we had full access to the dataset. To simulate the LBS access model, we implemented a $k$NN interface with Euclidean distance as the ranking function.

\noindent{\bf Real World Datasets:} In order to highlight the practicality of HDBSCAN, we evaluated it against a number of popular real-world LBS such as Yahoo! Flickr, Zillow, Redfin and Capital BikeShare. Note that, unlike the experiments on benchmark datasets, we do not know the ground truth cluster assignments. However, the cluster assignments produced by HDBSCAN had realistic clusters with valid real-world interpretations. 

The Yahoo! Flickr dataset contained almost 100 million images of which about 49 million are geotagged. In addition to image metadata such as title, description and other user tags, it also contains location information. We considered the subset of almost 118K images from Washington DC with location granularity of street level. We implemented a $k$NN interface over the dataset that returns the images geotagged with locations near the query location. We can consider locations with large number of geo-tagged images as clusters (e.g. popular tourist hotspots).

Zillow is a popular online real state website that helps users to find houses and apartments for sale/rent. It has a $k$NN interface that allows users to search for houses in a given location and filter search results based on attributes like price, area, home type etc. We crawled approximately 12K houses listed for sale in Dallas Fort Worth (DFW) area along with metadata such as house location, price, number of beds and baths, area in square feet etc. 

Redfin is an online real estate website similar to Zillow. The mobile application of Redfin provides a $k$NN interface where users can specify a location and get nearby houses sorted according to their distance from the query point.  We executed HDBSCAN over the Redfin house listings for the DFW area. 

Capital Bikeshare is a popular bicycle sharing system in Washington D.C. It has approximately 350 bike stations around D.C. They publish their rental data periodically every quarter. The dataset contains information about every rental such as start and end stations, rental date, duration etc. By combining this dataset with $k$NN interface provided by Google Places, we seek to cluster the stations based on the rental usage.

\noindent{\bf Performance Measures:} We measured the efficiency of HDBSCAN through query cost, i.e. the number of queries issued to LBS. The clustering quality is measured based on Rand index, Jaccard index and FolkesAndMallows index as defined in \S\ref{sec:datamodel}.

\noindent{\bf Parameter values:} Table~\ref{tbl:experimentParameterValues} shows the parameter values of HDBSCAN for each dataset. The value of $l$ was set to {\em minPts / 2}. The parameters for DBSCAN are set to the values that provided the best accuracy.

\begin{table}[!ht]
\small
\centering
\caption{Parameter values used in the experiments}
\label{tbl:experimentParameterValues}
\begin{tabular}{l |r | r |r | p{1cm}} 
  \hline
  {\bf Dataset}  	    &  {\bf $\epsilon$} 	& {\bf $minPts$} 	& { \# points} 	& {$k$} \\ \hline
  Chameleon t7.10k 	    & 20 			    & 14 			    & 10,000 		& 14 \\ 
  Chameleon t4.8k 	    & 20 			    & 19 			    & 8,000 		& 19 \\ 
  Chameleon t8.8k 	    & 15 			    & 6 			    & 8,000 		& 6 \\ 
  Zillow 		        & 0.02 			    & 10 			    & 12,250 		& 10 \\ 
  Yahoo Flickr DC	    & 0.0025 		    & 600 			    & 117,875 		& 600 \\ 
  Redfin 		        & 0.02 			    & 10 			    & NA 			& 200 \\ 
  Capital Bikeshare 	& 0.005 		    & 3 			    & NA 			& 20 \\ \hline
\end{tabular}
\end{table}

\subsection{Experiments over Benchmark Datasets}

\noindent {\bf Feasibility of HDBSCAN:} In our first set of experiments, we show that it is indeed possible to identify underlying cluster structures (even for complex shapes with varying sizes) using a restricted $k$NN query interface and limited sample size. Figure~\ref{fig:chamaleon_4_8k} visualizes the benchmark dataset Chameleon 4.8k while Figure~\ref{fig:chamaleon_4_8k_hdbscan} visualizes the output of HDBSCAN over the dataset (using only the sample points) with a query budget of $400$. To reduce clutter, we removed the noise points.  Figures~\ref{fig:chamaleon_8_8k} and \ref{fig:chamaleon_8_8k_hdbscan} show the results for Chameleon 8.8k. As the figures show, HDBSCAN followed by post-process merging of nearby clusters could identify the clusters using only the local view of the dataset.

\noindent {\bf Comparison with Baseline Algorithm:} Recall that a baseline approach for clustering over LBS data is to obtain sample using prior work such as \cite{liu2015aggrEst} and run DBSCAN over the sample. The unseen data points are assigned to the nearest cluster. If there are no cluster within a distance threshold, they are categorized as a noise. We compared the clustering quality of baseline algorithm and HDBSCAN with same query budget while varying the budget from 200 to 600. Figure~\ref{fig:ch_7_10_baseline} shows the results.  As expected, HDBSCAN outperforms baseline as the baseline often partitions the original clusters into many small clusters. Hence it is not a viable approach for tight query budgets. 

For the rest of the benchmark datasets experiments, we focus on the Chameleon 7.10k dataset (visualized in Figure~\ref{fig:ch7_10k_step4}). The results for other datasets were similar. Figure~\ref{fig:clusterQualityVsSFC} shows how the cluster quality varies when the three most popular space filling curves - Hilbert (HC), Peano (PC) and Z-curves (ZC) were used. Hilbert and Peano curves provide best cluster qualities. Hence, for the rest of the experiments we only consider HC and PC. Since our work is a best-effort implementation of DBSCAN over LBS, we treat the output of DBSCAN as ground truth. A cluster quality of $1.0$ is obtained when the cluster assignments of HDBSCAN and DBSCAN are identical.  

\noindent {\bf Cluster Quality versus Query Cost:}  In this experiment, we evaluate how the quality of clusters discovered by HDBSCAN is impacted when the query budget is varied. We vary query budget from 200 to 600 and Figure~\ref{fig:ch_clusterQualityVsQueryCost} shows the result. As expected, the cluster quality improves with higher query budget. Nevertheless, even for a query budget as small as $200$, the Rand index of HDBSCAN is $0.9$. Recall that Rand index measures the percentage of cluster assignments that are correct which highlights a 90\% accuracy of our algorithm.  

\noindent{\bf Varying $k$:}
The value of $k$ has a substantial impact on query cost. When the value of $k$ is higher than {\em minPts}, the additional results retrieved could be used to infer the density of neighboring grid cells. Figure~\ref{fig:ch_queryCostVsK} shows the result of experiments when the value of $k$ is varied. As expected, the query cost is reduced when the value of $k$ is increased. 

\noindent{\bf Varying Minimum Grid Size:} Next we demonstrate the effect of minimum grid size on query cost (Figure~\ref{fig:ch_queryCostVsMinGridSize}) and cluster quality (Figure~\ref{fig:ch_clusterQualityVsMinGridSize}). Intuitively, a large grid size reduces the total number of grids which in turn reduces the query cost. However, the grid size has a substantial impact on cluster quality. The relation between minimum grid size and cluster quality is not monotonic. If we set the minimum grid size too small the algorithm might partition the actual clusters into many smaller clusters. On the other hand, setting the grid size too large might merge the neighbor clusters in actual partitioning. In practice, our approach of adaptive grid sizes provides good results. 

\noindent{\bf Query Cost versus Node Partition Count:} At each level, standard Hilbert Curve partitions the existing cells into four equal size regions. However, in the adaptive space-filling curve approach we only partition a grid if required. We can divide the total query cost into two categories: i) Node splitting cost - queries that are issued on large grids that are partitioned later. ii) Leaf node query cost - queries executed at the leaf nodes of the tree. To reduce the node splitting cost, we can increase the number of children a node can be partitioned into. However, increasing this value may also increase the leaf node query cost. Figure~\ref{fig:ch_queryCostVsNodePartitionCount} illustrates how varying the number of partitions of a node impacts the query cost. When the number of partitions per node is small, the total query cost is high due to the higher node splitting cost. When number of partitions is increased, the leaf node query cost becomes higher. The optimal node partition count depends on the data distribution. When the distance between the clusters are large, small node partition count is better, whereas the opposite is true when clusters are close to each other. 

\noindent{\bf Query Cost versus Noise points (\%):}
In the final experiment, we investigate the impact of noise points on query cost. We varied the noise point count by randomly selecting a portion of noise points from the dataset. As expected, query cost increases with increase in the percentage of noise points. One of the major reasons for the increase in query cost is the impact the noise points have on the adaptive space filling curves. Without a well chosen threshold, the adaptive SFC algorithm could treat a sparse region (that would not have been explored if there is no noise) as a non-sparse region resulting in higher query cost.


\subsection{Experiments over Real-World LBS}

\begin{figure*}[!ht]
  \begin{minipage}[t]{0.48\linewidth}
    \centering
    \includegraphics[scale=0.4]{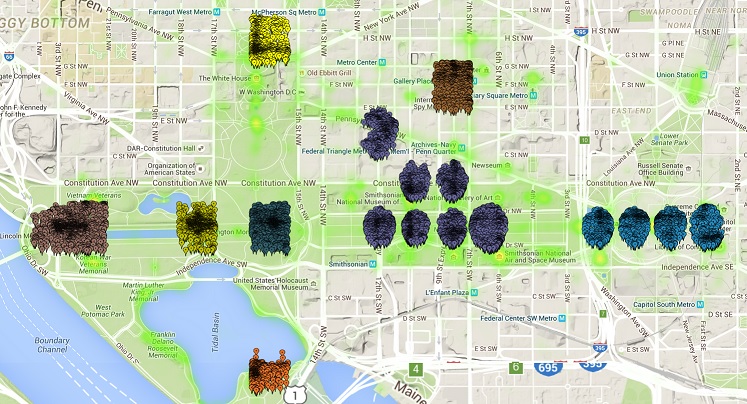}
    \caption{Clustering of Places in DC using Yahoo Flickr geotagged photo}
    \label{fig:yahooFlickrDC}
  \end{minipage}
  \hspace{1mm}
  \begin{minipage}[t]{0.48\linewidth}
    \centering
    \includegraphics[scale=0.4]{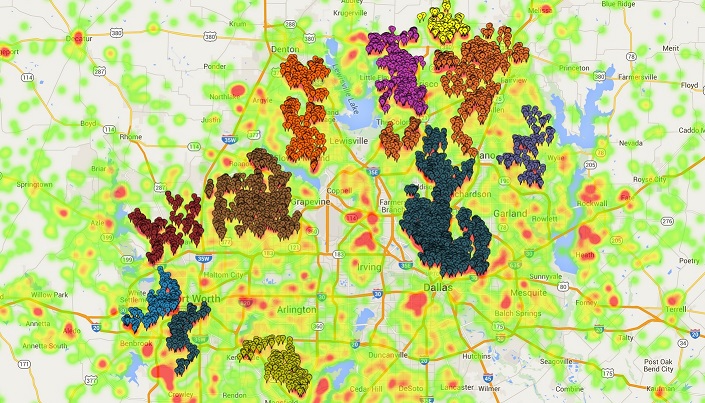}
    \caption{Clustering of houses in DFW area using Redfin}
    \label{fig:redfinDFW}
  \end{minipage}
\end{figure*}

\noindent{\bf Yahoo Flickr:} For the Yahoo Flickr dataset, we run HDBSCAN on almost 118K images geotagged with Washington, DC area. The output clusters are plotted over the heat map of photos in Washington, DC. Figure~\ref{fig:yahooFlickrDC} shows the comparison of discovered clusters with actual photo distribution. We can see that the clusters are located at the popular tourist spots of  Washington, DC. This was also corroborated by the analysis of most frequent user tags from image metadata within each cluster. We used the population count information of US census data as external knowledge for this experiment.

\begin{figure*}[!ht]
  \begin{minipage}[t]{0.48\linewidth}
    \centering
    \includegraphics[scale=0.4]{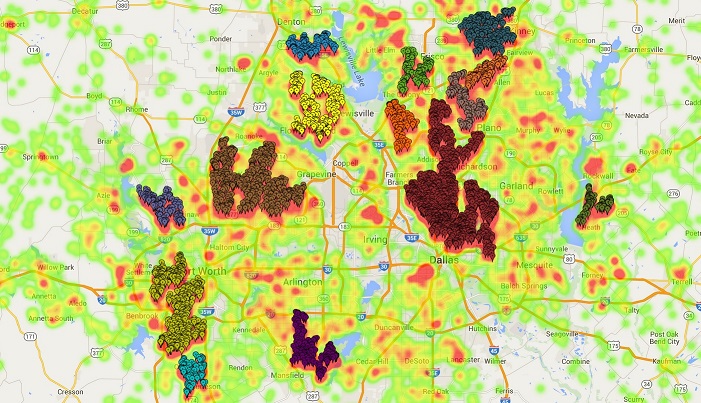}
    \caption{Clustering of houses in DFW area on Zillow dataset}
    \label{fig:zillowDFW}
  \end{minipage}
  \hspace{1mm}
  \begin{minipage}[t]{0.48\linewidth}
    \centering
    \includegraphics[scale=0.4]{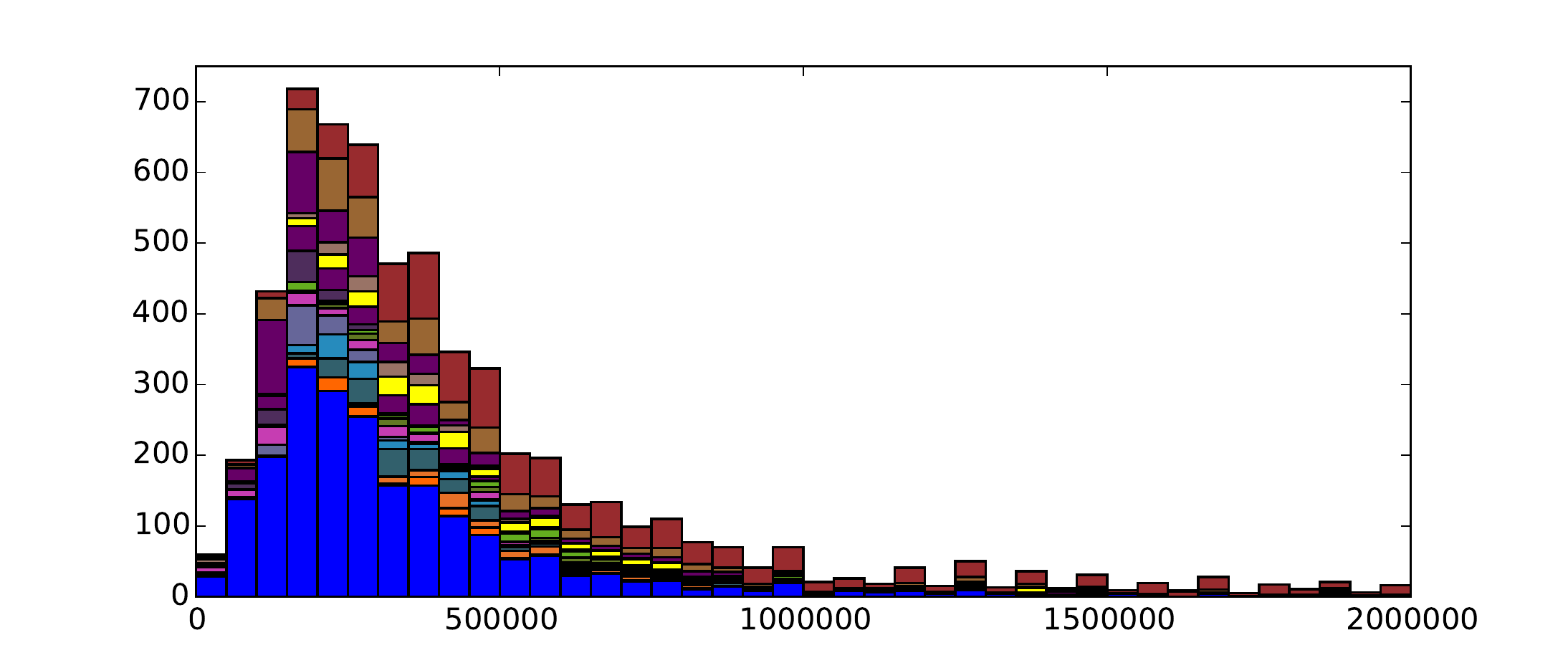}
    \caption{Price distribution of house for sales in DFW area}
    \label{fig:zillowDFWPriceDistribution}
  \end{minipage}
\end{figure*}

\noindent{\bf Zillow:} The results of running HDBSCAN over houses listed in DFW of Zillow website can be found in Figure~\ref{fig:zillowDFW}. In order to compare the discovered clusters with actual distribution of ``Houses for Sale'' in DFW area, we plot the houses inside each cluster over the heat map of houses in Zillow dataset. Only clusters containing at-least 100 houses are shown. We can see that HDBSCAN discovers clusters of different shapes that matches the actual distribution of house in DFW. Since we have access to full dataset, we also performed basic statistical analysis of the house prices in discovered clusters. As expected, clusters near Dallas and Forth-Worth downtown have higher average price compared to others. The price distribution of houses in DFW area is shown as histogram in Figure~\ref{fig:zillowDFWPriceDistribution}. For each price bin, the corresponding count in each cluster is shown with different color. Count of houses in clusters of size smaller than 100 are combined altogether and shown with blue bar. We can see that clusters near urban areas are at the right side of the histogram and clusters in rural area are placed the left side. We used the ``Housing Units'' information of US Census data as external knowledge for optimization \cite{uscensusdata}. 

\begin{figure*}[!ht]
  \begin{minipage}[t]{0.48\linewidth}
    \centering
    \includegraphics[scale=0.5]{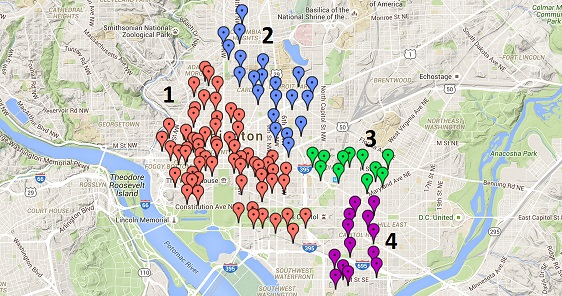}
    \caption{Clustering of Capital Bikeshare stations in DC}
    \label{fig:capitalBikeshare}
  \end{minipage}
  \hspace{1mm}
  \begin{minipage}[t]{0.48\linewidth}
    \centering
    \includegraphics[scale=0.4]{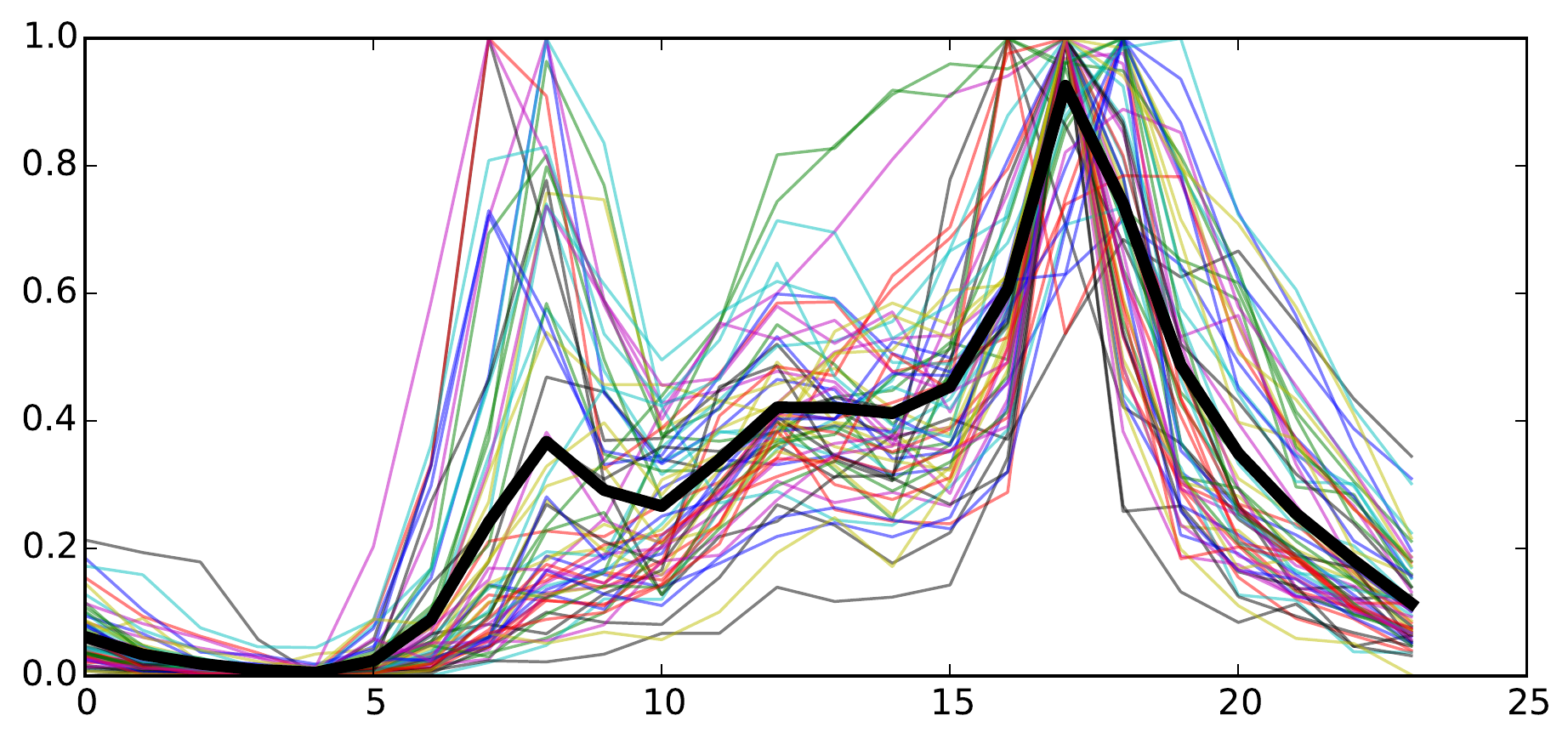}
    \caption{Average Rent per hour distribution for Cluster 1} 
    \label{fig:capitalBikeshare_rentFrequenceCluster}
  \end{minipage}
\end{figure*}


{\noindent \bf Redfin:}
We also performed clustering of houses in DFW area listed in Redfin website. This was an online experiment conducted live over Redfin. The output clusters are shown in Figure~\ref{fig:redfinDFW}. The list of identified clusters are similar to those identified from Zillow dataset. The minor differences are due to the fact that some houses were not listed in both services.

\vspace{1mm}
{\noindent \bf Capital Bikeshare:}
We used Google Places API to perform clustering on Capital Bikeshare (CB) stations in Washington D.C area. Since Google Maps does not have complete information about all the CB bike stations in D.C., we augmented the clustering results by adding the missing bike stations to its nearest cluster discovered using Google Place API. Figure~\ref{fig:capitalBikeshare} shows the clusters identified by HDBSCAN. There is one big cluster at the center of DC (shown in red marker) and three smaller clusters at the  outer side area(shown in blue and green markers). We validated the identified clusters from the historical rental data for each bike stations\footnote{https://www.capitalbikeshare.com/trip-history-data}. Figure~\ref{fig:capitalBikeshare_rentFrequenceCluster} displays the distribution of average rent count (normalized to 1) of each bike stations for Cluster 1 (red markers). We can see that stations in cluster 1 has a different hourly rental distribution from other stations. Specifically, we can see that there is a difference in the usage pattern among stations in CB. Bike stations close to the center of the city are more frequently rented at the evening time compared to other times of the day. Whereas stations located outside have two peaks in the morning and evening. Analysis of other clusters showed similar distinct usage pattern.

\section{Related Work}
\label{sec:relWork}

There has been extensive work on spatial data mining of which clustering is a major technique. 
Our problem is substantially different from traditional clustering problems as they have access to entire database. To the best our knowledge, our work is the first to propose enabling DBSCAN (or any clustering algorithm) over a LBS. We do not challenge or improve the cluster definitions over traditional databases, but rather enable the algorithmic design according to a given cluster definition over an LBS with a limited kNN interface. Please refer to \cite{berkhin2006survey} for a detailed survey of clustering algorithms. Prior work on finding cluster boundaries \cite{xia2006border,qiu2007brim} do not apply here as they also require access to entire database. 
Due to the increasing popularity of LBS, key problems such as sampling\cite{WHL14,LSW+12,liu2015aggrEst,DKA+11}, aggregate estimation\cite{liu2015aggrEst,DKA+11} and crawling\cite{YGZ16} over a restricted query interface such as $k$NN have been recently studied.  
SFC has been studied in the context of multi-dimensional indexing where the multi-dimensional data is mapped to one dimension, enabling simple and well understood one-dimensional indexing algorithms to be utilized \cite{mokbel2003analysis,moon2001analysis}. Theoretical analysis of the clustering properties of SFC has been studied in \cite{xu2014optimality,jagadish1990linear,moon2001analysis,mokbel2003analysis}.

\section{Final Remarks}
\label{sec:finalRemarks}

In this paper, we explore the problem of enabling DBSCAN over an LBS with only a limited, kNN query interface. 
We developed HDBSCAN that uses an adaptive space-filling curve algorithm as a subroutine to map 2D data to 1D, clusters the 1D data and then merges the resulting 1D clusters into eventual 2D clusters.
Extending our results to other density-based clustering algorithms is left for future work.
We verified the effectiveness of our algorithms by conducting comprehensive experiments on benchmark datasets and multiple real-world LBS. 

\bibliographystyle{abbrv}
\bibliography{references}

\end{document}